\documentclass[final]{IEEEtran}
\IEEEoverridecommandlockouts
\usepackage{color}

\usepackage{times}
\usepackage[final]{graphicx}
\usepackage{amsmath,amsfonts}
\usepackage{amsthm}
\usepackage{amssymb,amsbsy}

\usepackage{cite}
\usepackage{multirow}

\usepackage{algorithm}
\usepackage{algpseudocode}
\usepackage{verbatim}

\newcommand{\ignore}[1]{}
\newcommand{\hspp}{\hspace{0.05in} }

\newsavebox{\savepar}

\begin{document}
%\title{Spherical Coverage and Implications for Millimeter Wave Systems}
%\title{Antenna Placement and Implications for Spherical Coverage in Millimeter Wave
%Systems}
%\title{Tradeoffs Between Two Antenna Designs for Millimeter Wave Systems}
\title{Hand and Body Blockage Measurements with Form-Factor User Equipment at 28 GHz}
\author{\normalsize Vasanthan Raghavan, Sonsay Noimanivone, Sung Kil Rho, Bernie Farin, Patrick \\
Connor, Ricardo A.\ Motos, Yu-Chin Ou, Kobi Ravid,
M.\ Ali Tassoudji, Ozge H.\ Koymen, and Junyi Li
%\author{\normalsize Vasanthan Raghavan$^{\dagger}$,
%Mei-Li (Clara) Chi$^{\ddagger}$, M.\ Ali Tassoudji$^{\star}$, \\
%Ozge Hizir Koymen$^{\dagger}$, and Junyi Li$^{\dagger}$ \\
%$^{\dagger}$Qualcomm Flarion Technologies, Inc., Bridgewater, NJ, 08807, USA \\
%$^{\ddagger}$Formerly with Qualcomm Technologies, Inc., San Diego, CA, 92121, USA \\
%$^{\star}$Qualcomm Technologies, Inc., San Diego, CA, 92121, USA
}

\maketitle
%\IEEEpeerreviewmaketitle
\vspace{-10mm}

\begin{abstract}
\noindent
Blockage by the human hand/body is an important impairment in realizing practical millimeter wave
wireless systems. Prior works on blockage modeling are either based on theoretical studies of
double knife edge diffraction or its modifications, high-frequency simulations of electromagnetic
effects, or measurements with experimental millimeter wave prototypes. While such studies are
useful, they do not capture the form-factor constraints of user equipments (UEs). In this work,
we study the impact of hand/body blockage with a UE at $28$ GHz built on Qualcomm's millimeter
wave modem, antenna modules and beamforming solutions. We report five exhaustive and controlled
studies with different types of hand holdings/grips, antenna types, and with directional/narrow
beams. For both hard as well as loose hand grips, we report considerably lower blockage loss
estimates than prior works. Critical in estimating the loss is the definition of a ``region of
interest'' (RoI) around the UE where the impact of the hand/body is seen. Towards this goal, we
define a RoI that includes the spatial area where significant energy is seen in either the no
blockage or blockage modes. Our studies show that significant spatial area coverage improvement
can be seen with loose hand grip due to hand reflections.
%A simplistic and gross estimate of blockage loss is $8.5$-$17$ dB with a hard hand grip
%and $3.5$-$11$ dB with a loose hand grip, both of which are considerably lower than prior estimates
%for blockage loss. However, critical to estimating the loss is the definition of a ``region of
%interest'' (RoI) in the sphere around the UE where the impact of the hand/body is seen. Towards
%truly capturing the impact of the hand/body, we define a RoI that includes the spatial area where
%significant energy is seen in either the no blockage or blockage modes. Our studies show that
%while the hard hand grip mode does not result in any substantial hand reflections, significant
%spatial area coverage improvement can be seen in the loose hand grip mode due to hand reflections.
\end{abstract}


\section{Introduction}
Over the last ten years, the interest in millimeter wave carrier frequencies has
transformed from an academic/theoretical pursuit to commercial deployments. The first
wave of commercial form-factor user equipments (UEs) are already available in the market
with the physical layer operation conforming to the Third Generation Partnership
Project (3GPP) standard specifications in Release 15. Despite this essentially
mature
background~\cite{rusek,hur,sun,brady_tcom,oelayach,raghavan_jstsp,vasanth_jsac2017,roh,rangan,ghosh,vasanth_comm_mag_16}
in both the theory and practice of millimeter wave systems, there is still considerable
and growing interest in understanding the performance limits of such systems imposed by
the channel and propagation characteristics, radio frequency (RF) and hardware constraints
and their impact and implications for low-cost, low-complexity and power-efficient physical
layer design. The focus of this work is on one such impairment: blockage of millimeter wave
signals at the UE end due to human hand and body.

Given that blockage is not a dominant impairment at sub-$6$ GHz carrier frequencies, a number
of prior works have focussed on modeling blockage and understanding its implications on
millimeter wave system performance. In particular, wireless standardization efforts at $60$ GHz
for $802.11$(ad) WiFi systems use ray-tracing studies to propose a human blockage
model~\cite[Sections 3.3.8, 3.5.7, 5.3.9, 8]{802d11_maltsev}. This model reflects the
probability that a blockage event happens, a distribution for blockage loss
conditioned on it happening and time-scale modeling for blockage events. For
cellular millimeter wave systems, the 3GPP TR38.901~\cite[pp.\ 53-57]{3gpp_CM_rel14_38901}
proposes a flat $30$ dB loss over a defined blockage region for the UE in either the
Portrait or Landscape modes. The loss region is modeled using data from studies with a
form-factor experimental millimeter wave UE mock-up/prototype at $28$ and $60$ GHz and the
loss is motivated by a survey of measurement studies with human/body blockage. The Mobile and
wireless communications Enablers for the Twenty-twenty Information Society (METIS) project has
proposed a human blockage model based on the double knife edge diffraction (DKED) framework
in~\cite[pp.\ 39-41, 160-162]{metis2020_tcom}. Human blockage measurements over a wideband
setup at $60$ GHz has been considered in~\cite{peter_et_al}, where comparisons are made in
terms of model fitting with the DKED and the uniform theory of diffraction (UTD) frameworks.
Human blockage measurements using a $73$ GHz horn antenna setup is considered
in~\cite{maccartney_vtc2016,maccartney_2017,5G_whitepaper,rappaport_blockage2017,maccartney_2017_gcom}
and substantial losses ($30$-$40$ dB) are reported.

In terms of form-factor studies, the impact of blockage at $15$ GHz is studied
in~\cite{zhao_ericsson} and subarray diversity is recommended for overcoming the deleterious
effects of blockage. More recently, the loss with blockage is estimated in a prior work of
ours~\cite{vasanth_blockage_tap2018} using a $28$ GHz form-factor prototype performing
real-time beam switching/management and operating according to the system level specifications
analogous to the 3GPP framework albeit with a proprietary subframe structure. This study
reported an order-of-magnitude smaller blockage losses in beamformed systems than
prior modeling efforts (e.g., $30$ dB loss in TR38.901) and attributed the discrepancies
in loss due to beamwidth differences between commercial-quality phased arrays and horn
antennas (horn antennas have narrower beamwidths than UE-side phased arrays). In this
study, blockage loss was estimated with over-the-air (OTA) measurements of beamformed
received power differential between the no blockage and blockage scenarios. While such
studies are useful in understanding the practical impact of blockage, the received
power differential is a function of the channel environment (rich vs.\ non-rich multi-path
clusters) and the set of beam weights with which the link has been established between the
base-station and the UE (which determines the dominant cluster in the channel excited in
beamforming). Another important caveat common to all prior studies on blockage modeling
is that they are either based on ray-tracing or electromagnetic simulation studies, or
with experimental prototypes that may/may not be a form-factor implementation.

\noindent {\bf \em \underline{Contributions:}} In this context, this work reports blockage
losses with a commercial form-factor UE
%(which we denote as the ``Qualcomm Reference Design'' or QRD)
operating at $28$ GHz. The UE is equipped with a commercial grade millimeter wave
modem, %(Qualcomm's X50 modem~\cite{qcom_x50}),
antenna module solution %(QTM052 modules~\cite{qcom_qtm052})
and driven by a beam management software solution that adheres to
the 3GPP system level protocol specifications in Rel.~15~\cite{3gpp_38series}. The
antenna module incorporated here uses a $4 \times 1$ dual-polarized patch array and two
$2 \times 1$ dipole arrays across two polarizations/layers. Multiple commercial millimeter
wave UEs available in the market today use similar antenna, modem and system level software
implementations that realize low-overhead and low-complexity analog/RF beamforming and beam
tracking~\cite{raghavan_jstsp}, thereby improving signal range and coverage. Thus, this work is
directly relevant in understanding blockage from a practical/implementation perspective.

In contrast to OTA measurements of~\cite{vasanth_blockage_tap2018}, we report five
{\em controlled} studies in an anechoic chamber that allows us to understand blockage by
studying beam patterns over a sphere with Freespace/no blockage, a hand
phantom %~\cite{speag_hand_phantom}
used to emulate blockage, and a real/true human
holding the phone with hand and body of the human blocking the signals. By studying the
beam patterns over a sphere, the impact of the channel used to establish a beamformed
link is removed and we can showcase the true impact of blockage in different directions.
The reported studies correspond to different targeted antenna arrays of different
dimensions ($4 \times 1$ patch array vs.\ $2 \times 1$ dipole array), different UE
orientations (Portrait vs.\ Landscape), and different hand holdings/grips. The grips
studied here include a ``hard'' hand holding grip where the hand completely engulfs all
the antenna elements in the array with minimal air gaps between the fingers, a ``loose''
hand holding grip where only a few fingers engulf some of the antenna elements in the
array with the remaining antenna elements seeing unobstructed signals, and an
``intermediate'' hand holding grip where a few fingers engulf some antenna elements
with a big air gap between the palm of the hand and the remaining antenna elements.

\begin{table*}[htb!]
\caption{Summary Statistics From the Five Blockage Studies}
\label{table_brief_experiments}
\begin{center}
\begin{tabular}{|c||c|c|c|c|c||c|}
\hline
\hline
Study & Subarray type & UE orientation & Hand grip &
Gross loss & Relative $\%$ of & Relative RoI \\
& & & & estimate (in dB) & sphere lost %(in \%)
& improvement (in \%)
\\
\hline
\hline
1 & $4 \times 1$ patch & Portrait & Hard &
$8.6$ to $17.2$ & $66.7\%$ to $85.3\%$ & $2.1\%$ to $4.6\%$
\\ \hline
2 & $4 \times 1$ patch & Portrait & Loose & $3.6$ to $10.6$ &
$36.0\%$ to $43.9\%$ & $6.7\%$ to $8.0\%$
\\ \hline
3 & $2 \times 1$ dipole & Portrait & Hard & $15.9$ to $19.7$ &
$90.8\%$ to $100.0\%$ & $0\%$ to $1.7\%$
\\ \hline
4 & $2 \times 1$ dipole & Portrait & Loose & $0.4$ to $10.8$ &
$20.0\%$ to $25.7\%$ & $8.5\%$ to $12.9\%$
\\ \hline
5 & $4 \times 1$ patch & Landscape & Intermediate & $9.5$ to $12.7$ &
$68.6\%$ to $99.9\%$ & $8.9\%$ to $15.7\%$
\\
\hline
\hline
\end{tabular}
\end{center}
\end{table*}
%\end{center}

From our studies, we note the following broad conclusions.
\begin{itemize}
\item
We start with a gross estimate of blockage losses obtained by comparing the cumulative
distribution functions (CDFs) of radiated signal power with Freespace/no blockage and with
blockage. These gross loss estimates range from $8.5$-$17$ dB in the hard hand grip mode to
$3.5$-$11$ dB in the loose hand grip mode for the $4 \times 1$ subarray. These loss
estimates are {\em significantly}
lower than loss estimates at 3GPP~\cite{3gpp_CM_rel14_38901} and in prior
studies~\cite{maccartney_2017_gcom}. The estimates provided here are also consistent with
(and similar to) our prior work~\cite{vasanth_blockage_tap2018} that used a $28$ GHz
experimental prototype and estimated a mean loss of $15.3$ dB and $8.5$ dB for hand and
body blockage losses.

\item
Going further from all the prior works, we show that depending on the antenna type (dipole
or patch), array size ($4 \times 1$ vs.\ $2 \times 1$), type of beam used (scan angle and
beamwidth), material property of UE, and the user's hand properties (such as
hand grip, hand size, skin properties), etc.,
\begin{enumerate}
\item The hand can attenuate signals in a certain set of directions;
\item the hand can reflect energy in some set of directions; and
\item the hand can leave signal energy essentially unchanged in the
remaining set of directions.
\end{enumerate}
With a primary focus on blockage loss in prior works, the impact of hand reflections has
{\em \underline{not}} been
explored. In this context, this work provides a first understanding of this aspect.

\item
The impact of hand/body in terms of signal deterioration varies with direction. Thus, to
estimate the statistics of blockage loss, we need to define a ``region of interest'' (RoI)
where blockage is observed. We begin with a RoI (denoted as ${\cal R}_1$) corresponding to the
Freespace/no blockage region that is within a fixed signal threshold of its peak value. This
RoI does not capture hand reflections into regions which had a poor signal strength in the no blockage
mode. Thus, to consider the impact of hand reflections, we augment the above RoI (denoted as
${\cal R}_5$) with the region where signal strength in the blockage mode is also above a signal
strength threshold. A number of other intermediate RoIs (denoted as ${\cal R}_2$, ${\cal R}_3$
and ${\cal R}_4$) are also defined corresponding to different nuanced aspects of blockage.

\item
We show that ${\cal R}_1$ is sufficient to capture the impact of blockage in scenarios with a hard
hand grip where there are no prominent air gaps between fingers and hence there are minimal hand
reflections. On the other hand, ${\cal R}_5$ is necessary to understand blockage in scenarios with
a loose or intermediate hand grip, where a few antenna elements are unobstructed, or where a
significant air gap can be seen between some fingers and the antenna elements leading to hand
reflection gains.

\item
In general, studies with the hand phantom seem to be poorly correlated with true hand/body blockage
performance. This may be attributed to the need for careful tuning of the hand phantom as
well as its orientation and placement to correspond to how a real hand holding works.
Further, hand phantoms do not capture body effects.
Differences in electromagnetic behavior induced by material property differences between the
phantom material and the true human hand at millimeter wave carrier frequencies could also
account for discrepancies in the hand phantom performance relative to the real hand. The use of
hand phantoms requires more extensive studies in the future.

\item
The summary statistics in terms of blockage performance for these five studies are briefly
described in Table~\ref{table_brief_experiments}. These summary statistics illustrate the
gross loss estimate (in dB) with blockage across different spherical coverage levels,
relative fraction of the sphere lost at different effective isotropically radiated power (EIRP)
levels, and how hand reflections and the new definition of ${\cal R}_5$ relatively
improves the RoI/spherical coverage (over ${\cal R}_1$).
\end{itemize}

This paper is organized as follows. Section~\ref{sec2} explains the experimental setup
considered in this work in terms of hand/body blockage measurements. Section~\ref{sec3}
considers the $4 \times 1$ patch subarray with a hard hand grip and studies the impact
of hand/body blockage in careful detail by exploring the need for different RoIs in
understanding the implications of blockage. Section~\ref{sec4} performs similar studies
for the four other scenarios considered in this work. Section~\ref{sec5} develops
models for blockage for all the five scenarios considered here as well as compares
the physical layer implications of this work with prior models on blockage.
Section~\ref{sec6} concludes the paper.

\section{Experimental Setup}
\label{sec2}

We now explain the experimental setup used for measuring hand/body blockage in this
paper.

\subsection{User Equipment}
\begin{figure*}[htb!]
\begin{center}
\begin{tabular}{cc}
\includegraphics[height=2.5in,width=2.8in]{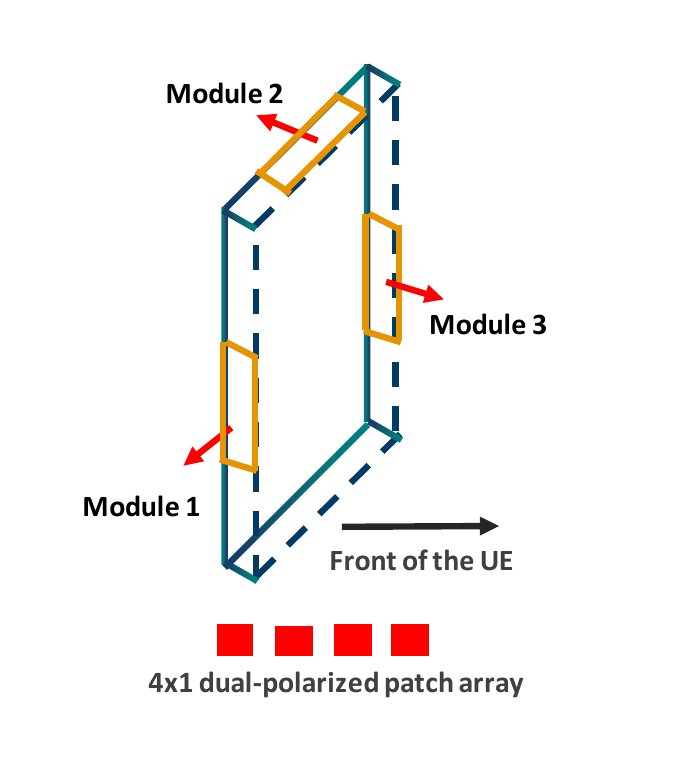}
&
\includegraphics[height=1.4in,width=3.4in]{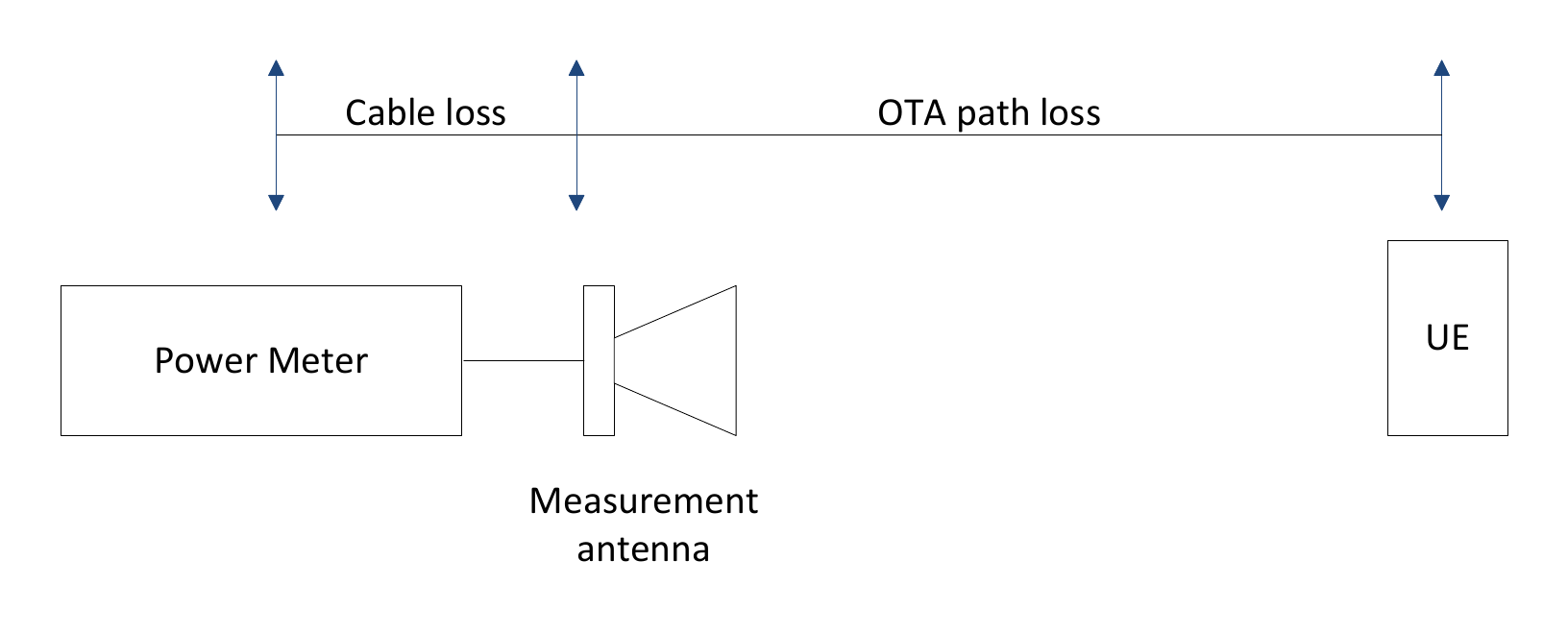}
\\
(a) & (b)
%\\
%\includegraphics[height=2.3in,width=3.0in]{data_plots_blockage/fig_paper_Fig1_Freespace_beam2_AG0Patch_Mod0.eps}
%&
%\includegraphics[height=2.3in,width=3.0in]{data_plots_blockage/fig_paper_Fig1_Freespace_beam3_AG0Patch_Mod0.eps}
%\\
%(c) & (d)
\end{tabular}
\caption{\label{fig_UE_Edge_Design}
(a) Pictorial illustration of the relative positions of the three antenna modules
in the UE. (b) Pictorial illustration of the measurement setup. }
\end{center}
%\vspace{-5mm}
\end{figure*}

%\noindent {\bf \em \underline{User Equipment:}}
The UE used in this study is %the ``Qualcomm Reference Design'' (QRD) which is
equipped with a millimeter wave modem %~\cite{qcom_x50}
operating at $28$ GHz and using a 3GPP Release 15 spec-compliant software solution that performs
intelligent beamforming and beam tracking. From an antenna module perspective, the UE consists
of three modules %~\cite{qcom_qtm052}
denoted as Modules 1-3. These modules are equipped on
the three edges/sides (two long edges and the top short edge) of the UE. Each %QTM052
antenna module has a
$4 \times 1$ dual-polarized patch array as well as two $2 \times 1$ dipole arrays that allows
dual-polarized transmissions via two RF chains at $28$ GHz. See Fig.~\ref{fig_UE_Edge_Design}(a)
for an illustration of the UE with the locations of the three antenna modules and the antenna setup within
each module. Since the UE is a pre-commercial design, it has a width beyond $72$ mm making it
a wide-body phone design.
%The QRD's size is $75$ mm $\times 158$ mm $\times 8$ mm, which puts
%it in the category of wide-body phone designs.

\begin{figure*}[htb!]
\begin{center}
\begin{tabular}{ccc}
\includegraphics[height=1.7in,width=1.7in]{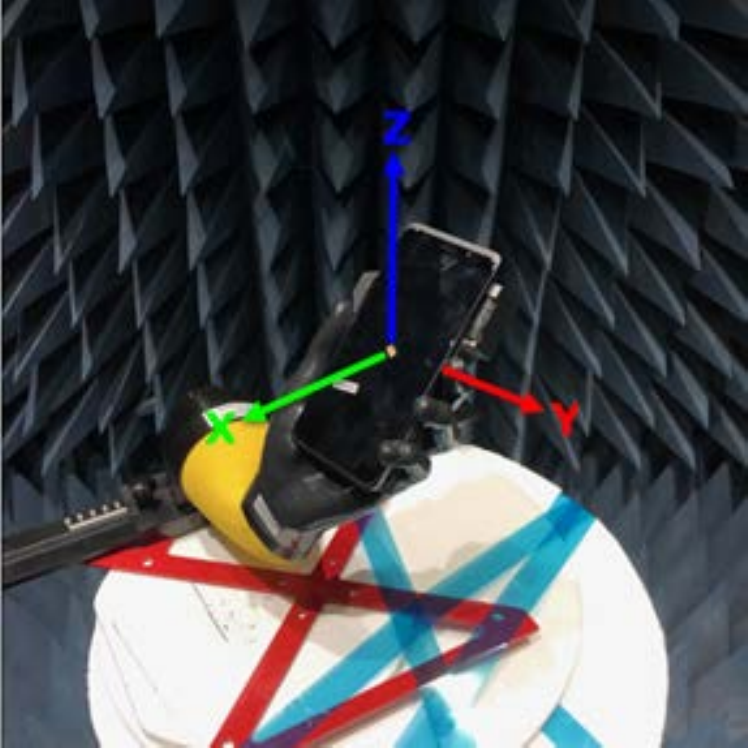}
&
\includegraphics[height=1.7in,width=1.7in]{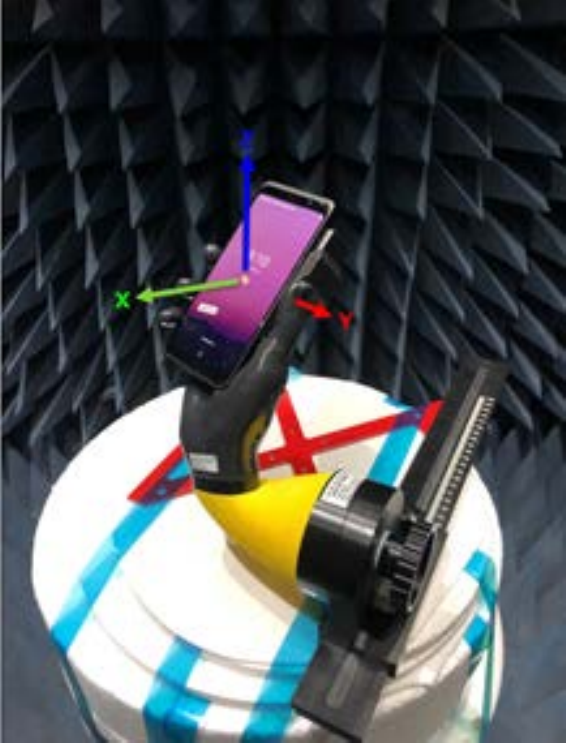}
&
\includegraphics[height=1.7in,width=2.5in]{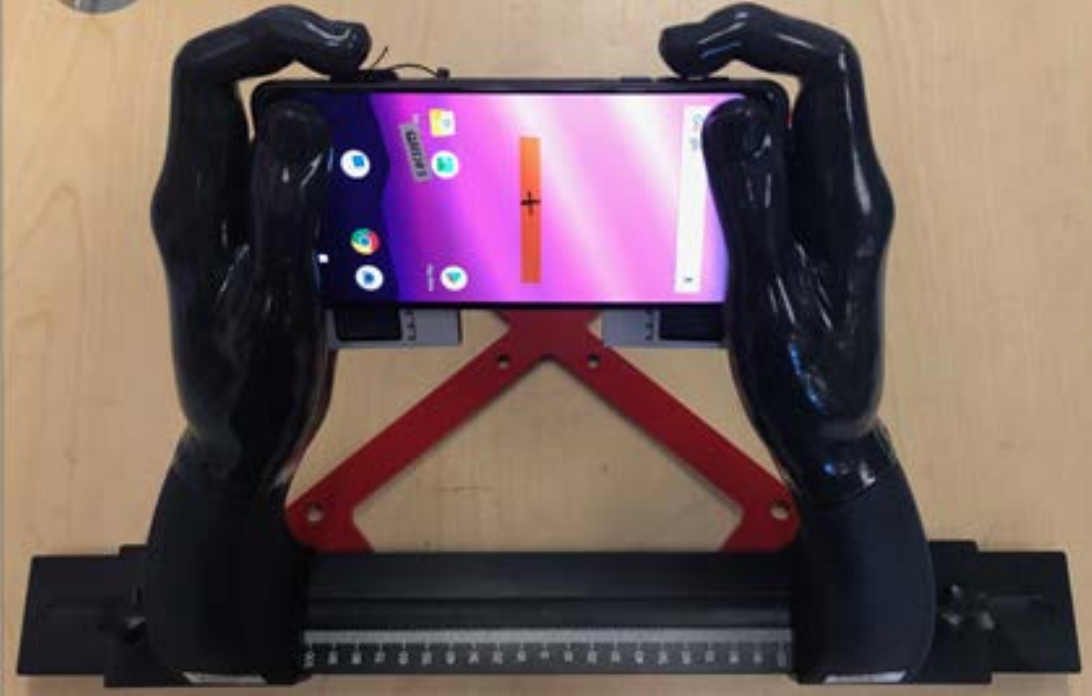}
\\
(a) & (b) & (c)
\\
%%%%%%%%%%%%%%%%
\includegraphics[height=1.9in,width=1.7in]{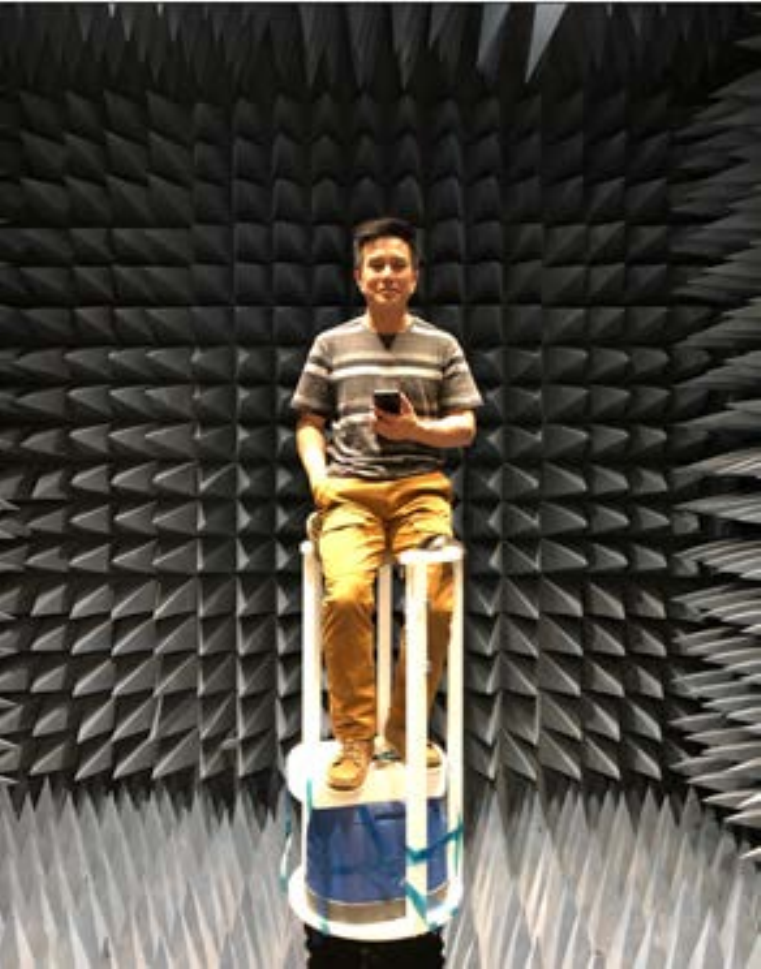}
&
\includegraphics[height=1.9in,width=1.7in]{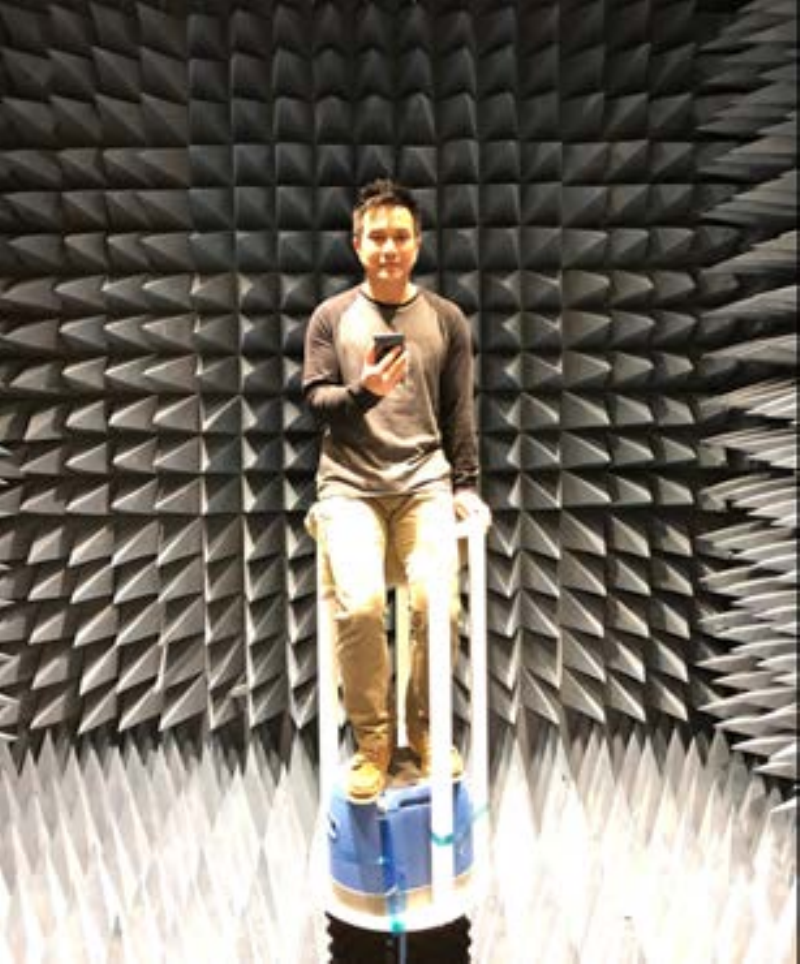}
&
\includegraphics[height=1.9in,width=1.7in]{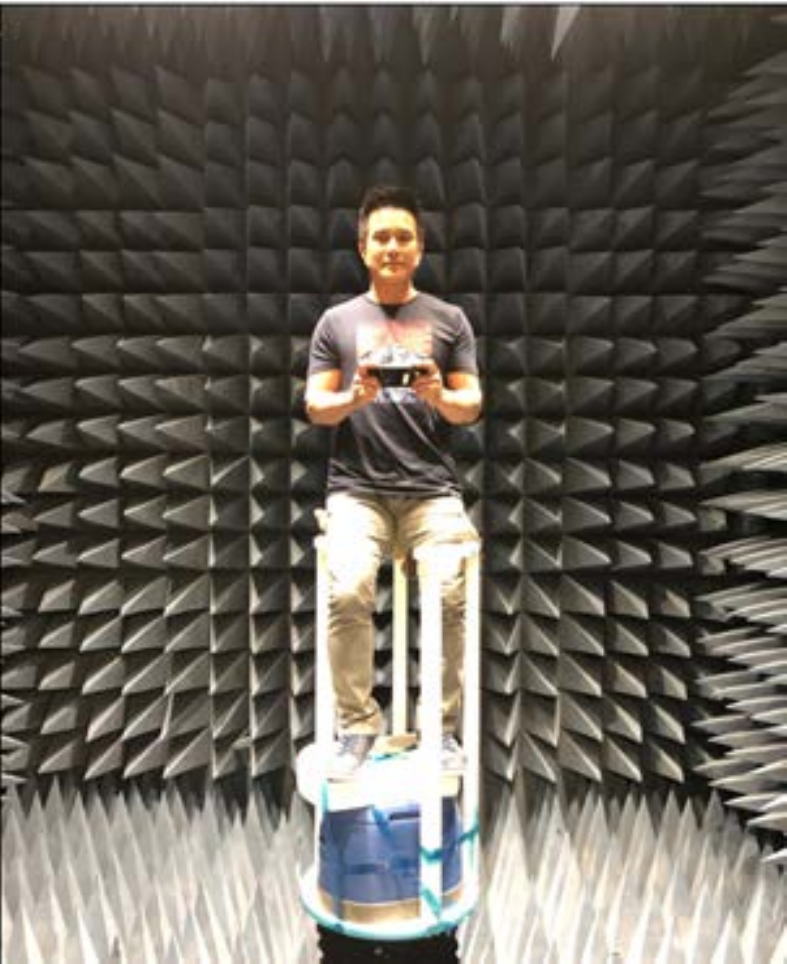}
\\
(d) & (e) & (f)
\end{tabular}
\caption{\label{fig_hand_phantoms}
Illustration of hand phantom and a true hand placed on the left hand side of the UE
((a) and (d)) and right hand side of the UE ((b) and (e)), both in Portrait mode.
(c) and (f): Hand phantom and true hand in gaming/Landscape mode.}
\end{center}
%\vspace{-5mm}
\end{figure*}

\subsection{Chamber Measurement Setup}
%\noindent {\bf \em \underline{Chamber Measurement Setup:}}
The anechoic chamber setup
is pictorially illustrated in Fig.~\ref{fig_UE_Edge_Design}(b) and is now described. The
measurement (receiving) antenna is an off-the-shelf dual-polarized broadband horn
antenna (covering $18$-$40$ GHz) %from A-Info\footnote{Model no. LB-SJ-180400, See
%{\tt http://www.ainfoinc.com/en/}.} %p\_ant\_h\_dual.asp}.}
with an antenna gain of
$\approx 14$ dBi at $28$ GHz. %and $\approx 16$ dBi at $38.5$ GHz.
The $3$ dB beamwidth in the H and E planes of the horn antenna at $28$ GHz are $35.8^{\sf o}$ and
$31.9^{\sf o}$, respectively. %, and $28.6^{\sf o}$ and $21.3^{\sf o}$ at 38.5 GHz.
The UE is placed on a fiberglass pedestal in the center of the chamber. The transmit
power used with the active millimeter wave antenna module is $4$ dBm, which is well
within the 3GPP EIRP regulations for
commercial millimeter wave devices and is intended for short distance coverage
between the UE and the measurement antenna. The distance between the UE and the
measurement antenna is $\approx 1.50$ meters ($59$ inches).

Short RF flex cables (with some loss) are used to connect the measurement antenna with a %Rohde-Schwarz
power meter. The power level observed by the power meter ($P_{\sf rx}$) can be written as
\begin{eqnarray}
{P}_{\sf rx} = {\sf EIRP}_{\sf tx} + G_{\sf rx} - {\sf Path \hspp loss} - {\sf Loss}_{\sf cable},
%\underbrace{ G_{\sf rx} - {\sf Path \hspp loss} - {\sf Loss}_{\sf cable} }_{\sf OTA \hspp
%calibration},
\nonumber
\end{eqnarray}
where ${\sf EIRP}_{\sf tx}$ is the transmitted EIRP with reference plane set to the outer
surface of the back cover\footnote{The considered reference plane implies that the loss
due to the radome/back cover is lumped with the measured data and thus, it is not necessary
to worry about the angle-dependent radome loss.} %Further, the blockage scenario considered
%is the scenario where the finger/hand physically touches and/or blocks the radiating aperture
%even though the energy is actually transmitted by the QTM052 module that is hidden behind
%the back cover.}
of the UE side, $G_{\sf rx}$ is the gain of the measurement antenna,
and ${\sf Path \hspp loss}$ and ${\sf Loss}_{\sf cable}$ correspond to loss in OTA transmissions
(with a path loss exponent of $2$ since a line-of-sight path is maintained between the UE and the measurement
antenna) and loss in the cables connecting the horn antenna with the power meter, respectively.
To measure the received power level accurately, an OTA path loss calibration procedure is
performed to capture the impact of measurement antenna gain, cable loss and path loss. Note
that while ${\sf EIRP}_{\sf tx}$ can be estimated theoretically, measurements are needed to
understand the over-/under-estimation of single antenna elemental gains and the array gains
due to different sets of beam weights in different directions, as well as non-idealities in
the array geometry and impact of UE material properties on the observed beam patterns.

As the UE transmits with a certain set of beam weights, the measurement antenna is rotated
(control for the rotation is driven by an automated software) at $\approx 5^{\sf o}$ steps
in azimuth and elevation. Due to the design of the chamber, a limited $170^{\sf o}$ (of the
possible $180^{\sf o}$) coverage in elevation is possible leading to a coverage map of the
beam pattern over the $360^{\sf o} \times 170^{\sf o}$ part of the sphere. By carefully
choosing the UE orientation in Freespace and with blockage, the impact of the missing
$10^{\sf o}$ in elevation plane on the conclusions of this work can be made minimal. In
the tests conducted in this paper, the UE orientation in the testing positions means that
the missing $10^{\sf o}$ is towards the bottom of the UE which has no antenna module
coverage and is covered by side lobes of other antenna modules. Since the EIRP in these
side lobe directions is expected to be low, lack of measurements in the missing $10^{\sf o}$
is expected to affect the tail of the performance curves, which do not carry any major
impact on understanding the implications of blockage.

\subsection{Hand Phantom and True Hand Holding}
%\noindent {\bf \em \underline{Hand Phantom and True Hand Holding:}}
In the Portrait mode, we
first use an anthropomorphic hand phantom %from SPEAG (Model\footnote{See
%{\tt https://speag.swiss/products/em-phantoms/phantoms/ \newline
%sho-v3rw-c-lw-c/}.} no.\ SHO3TO110-V3RW-C/LW-C), %~\cite{speag_hand_phantom})
which is specifically designed for evaluating and optimizing OTA performance of ultra-wide mobile phone
devices (defined as having a width between $72$ and $92$ mm). The hand phantom is manufactured using a
silicone-carbon-based mixture with material properties conforming to the Cellular Telecommunications
Industry Association (CTIA) definitions and standards for hand phantoms. The use of a special
low-loss silicone coating %applied to SHO3TO110-V3RW-C/LW-C
extends its useable frequency range from
$3$ GHz to $110$ GHz. A low-loss high-precision data mode fixture %\footnote{See
%{\tt https://speag.swiss/products/em-phantoms/accessories/sho-fx-hotv4/}.}
%(SHO-FX-HOTV4)
is used for accurate
and stable positioning of the hand phantoms in the correct position for the testing of devices. In
the Landscape mode, we use %SPEAG's SHO3TO110-V3RTHG/LTHG model %\footnote{See {\tt https://speag.swiss/products/em-phantoms/phantoms/sho-v3rthg-lthg/}.}
%which is
a two-handed grip for
gaming mode studies in conjunction with a wrist extension %\footnote{See {\tt https://speag.swiss/products/em-phantoms/accessories/sho-rthg-lthg-dwv4/}.}
%(SHO-RTHG/LTHG-DWV4)
and the data mode fixture. %(SHO-FX-HOTV4).
The hand phantoms on the left and right edges of the UE in Portrait mode
and the hand phantom in the gaming/Landscape mode are illustrated in Figs.~\ref{fig_hand_phantoms}(a)-(c),
respectively.

For the true hand holding tests, a testing person holds the UE and sits in a static position
in the chamber while the measurements are conducted. Each test (scan over the sphere) takes
approximately $18$-$21$ minutes and thus different testing persons are employed in
the studies in this paper. The testing persons vary from having a small body size ($145$ lbs,
$5$ feet and $4$ inches and $165$ lbs, $5$ feet and $1$ inch) to a large body size ($214$ lbs,
$6$ feet and $1$ inch). Our studies show that while the blockage losses reported in this paper
have some minor dependencies on the size of the hand, palm and
fingers as well as the skin properties of the hand, these dependencies are secondary and
minor in comparison with the type of hand holding/grip %(hard, intermediate or loose)
relative to the antenna array dimensions involved and the steering direction of the beams
used.

In terms of hand holding, three broad categories of tests are identified:
\begin{itemize}
\item A hard hand grip with the right hand in the Portrait mode that completely
engulfs all the antenna elements in Module 3 with minimal air gaps between the fingers.
\item A loose hand grip with the left hand in the Portrait mode where only a few fingers
engulf some of the antenna elements in Module 3 with unobstructed signals from other
antenna elements.
\item An intermediate hand grip with two hands in the Landscape mode where a few fingers
engulf some antenna elements in Module 2 with a big air gap between the palm of the hand
and the remaining antenna elements.
\end{itemize}
These three hand grips are reflective of the testing person's holding
of the UE as illustrated in Figs.~\ref{fig_hand_phantoms}(d)-(f).

\section{Study 1: $4 \times 1$ Patch Subarray with a Hard Hand Grip} %with Analog Codebooks}
\label{sec3}

The deployment of multiple antennas at millimeter wave carrier frequencies can be leveraged to
improve the link margin via beamforming. Since a limited number of RF chains are available at the
UE end at millimeter wave carrier frequencies (the UE considered in this work has two RF chains,
which are used for
polarization-based transmissions), increased array gain is realized with analog/RF beamforming.
Here, a three-bit phase shifter and a variable gain amplitude control are used at each antenna element
to co-phase the signals along desired/pre-specified directions. A beamforming scheme realized with
a finite-sized analog/RF beam codebook of beam weights that steers energy along the dominant
cluster(s) in the channel is a good low-complexity near-optimal solution relative to the optimal
beamforming scheme (performing maximum ratio combining)~\cite{raghavan_jstsp}. The performance of
this codebook improves as the codebook size increases and approaches the performance of a
directional beamforming scheme with perfect knowledge of the dominant cluster in the channel
as seen at the base-station and UE ends~\cite{vasanth_gcom15}.
In this section, we study the beamforming performance of the analog codebook of beams (without/with
blockage) %for different subarrays. We start with
for the $4 \times 1$ patch subarray in Module 3 with a hard hand grip. Other subarrays and hand holdings
are considered in Sec.~\ref{sec4}. %the next section.
%the $4 \times 1$ patch and $2 \times 1$ dipole subarrays considered in this work.
%\subsection{Freespace Behavior}

\begin{figure*}[htb!]
\begin{center}
\begin{tabular}{ccc}
\includegraphics[height=2.0in,width=2.2in]{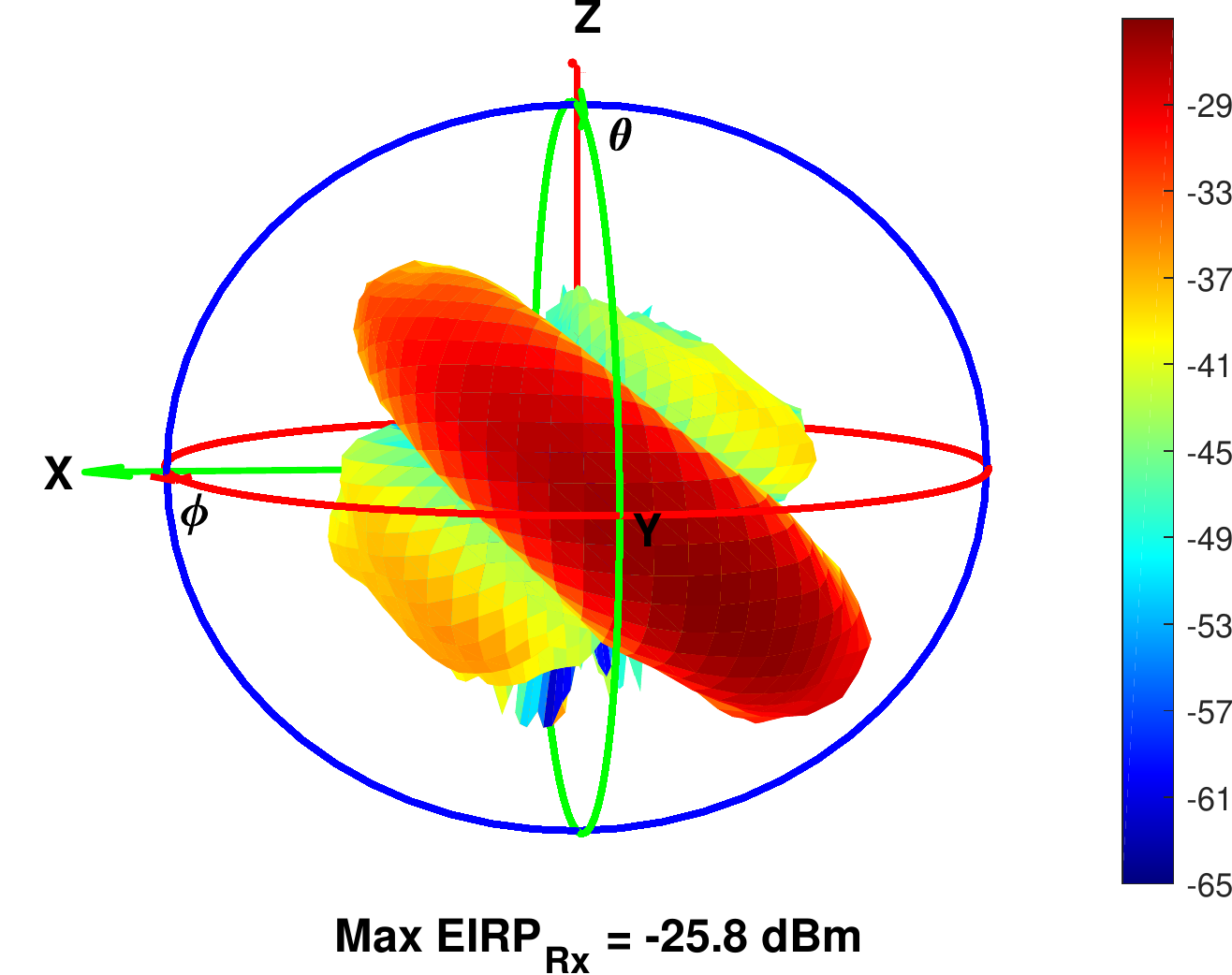}
&
\includegraphics[height=2.0in,width=2.2in]{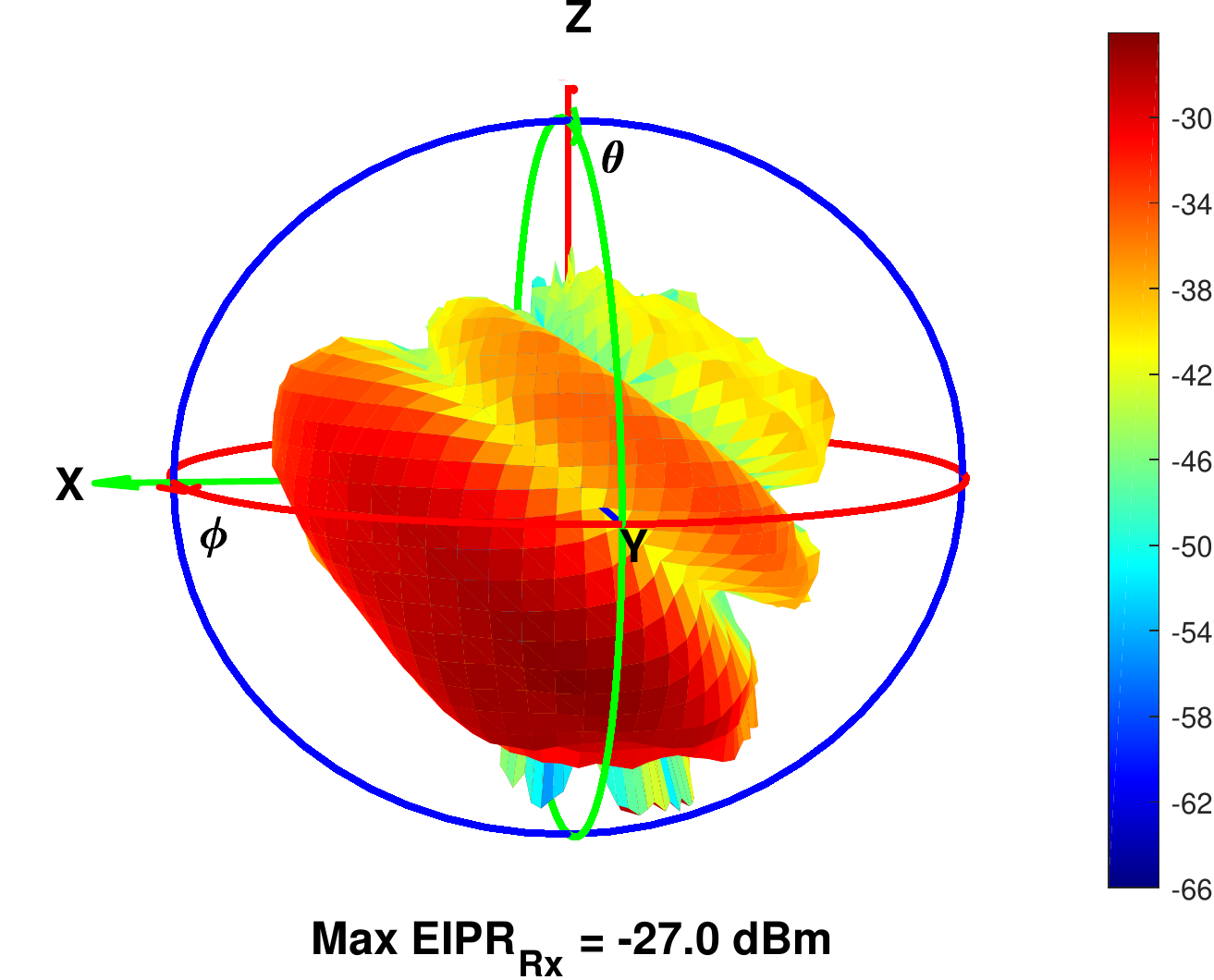}
&
\includegraphics[height=2.0in,width=2.2in]{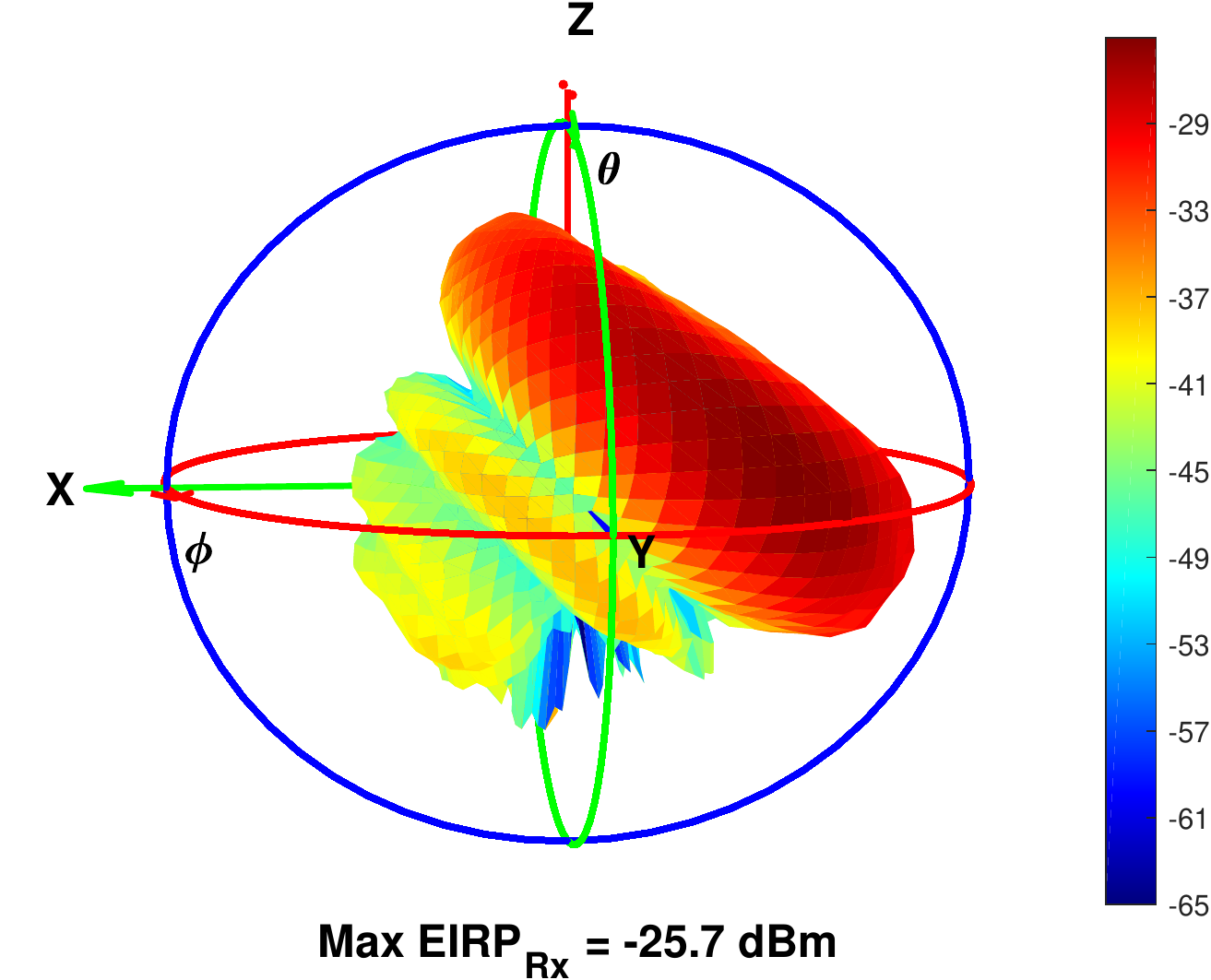}
\\
(a) & (b) & (c)
\\
\includegraphics[height=2.0in,width=2.2in]{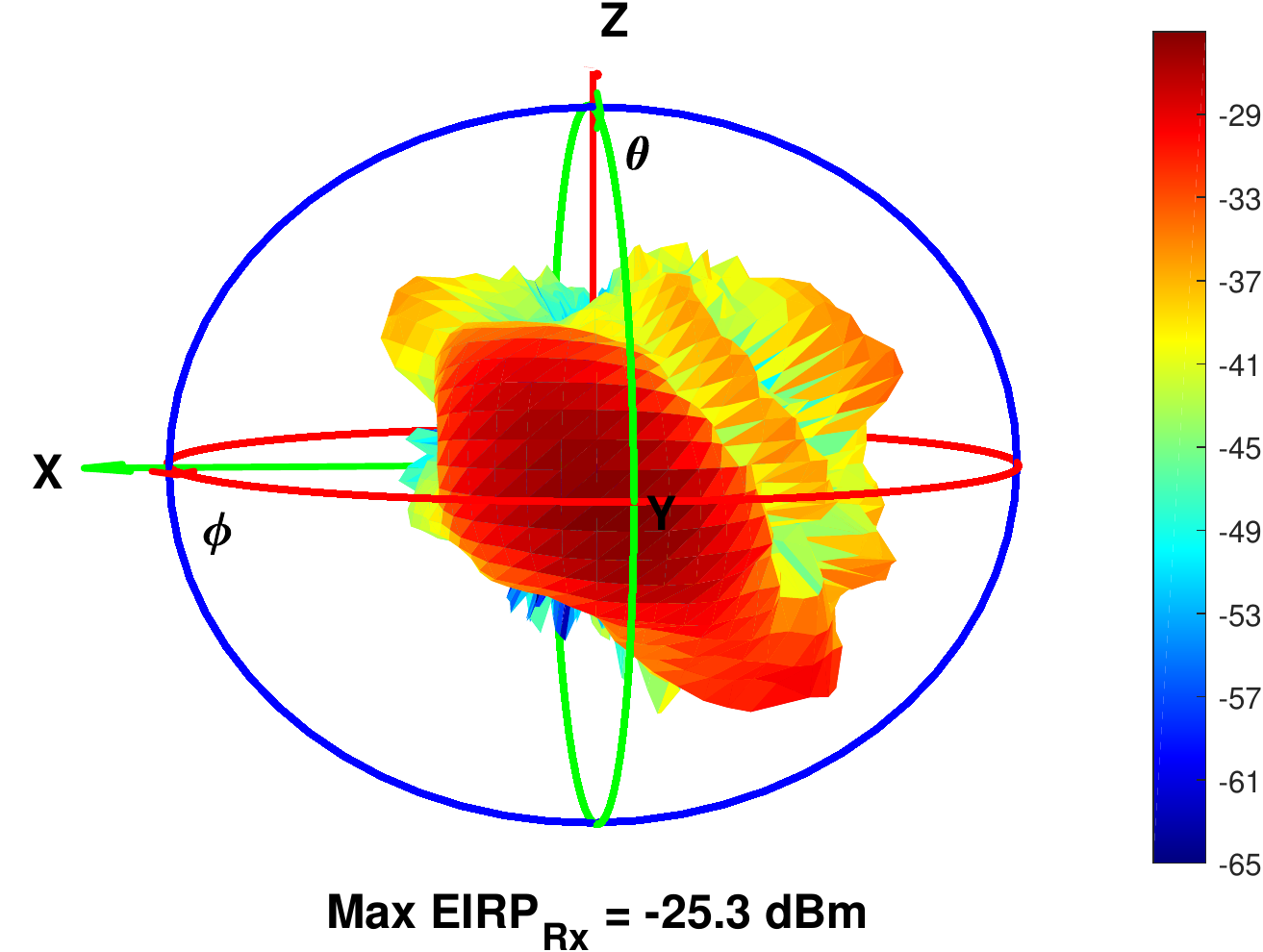}
&
\includegraphics[height=2.0in,width=2.2in]{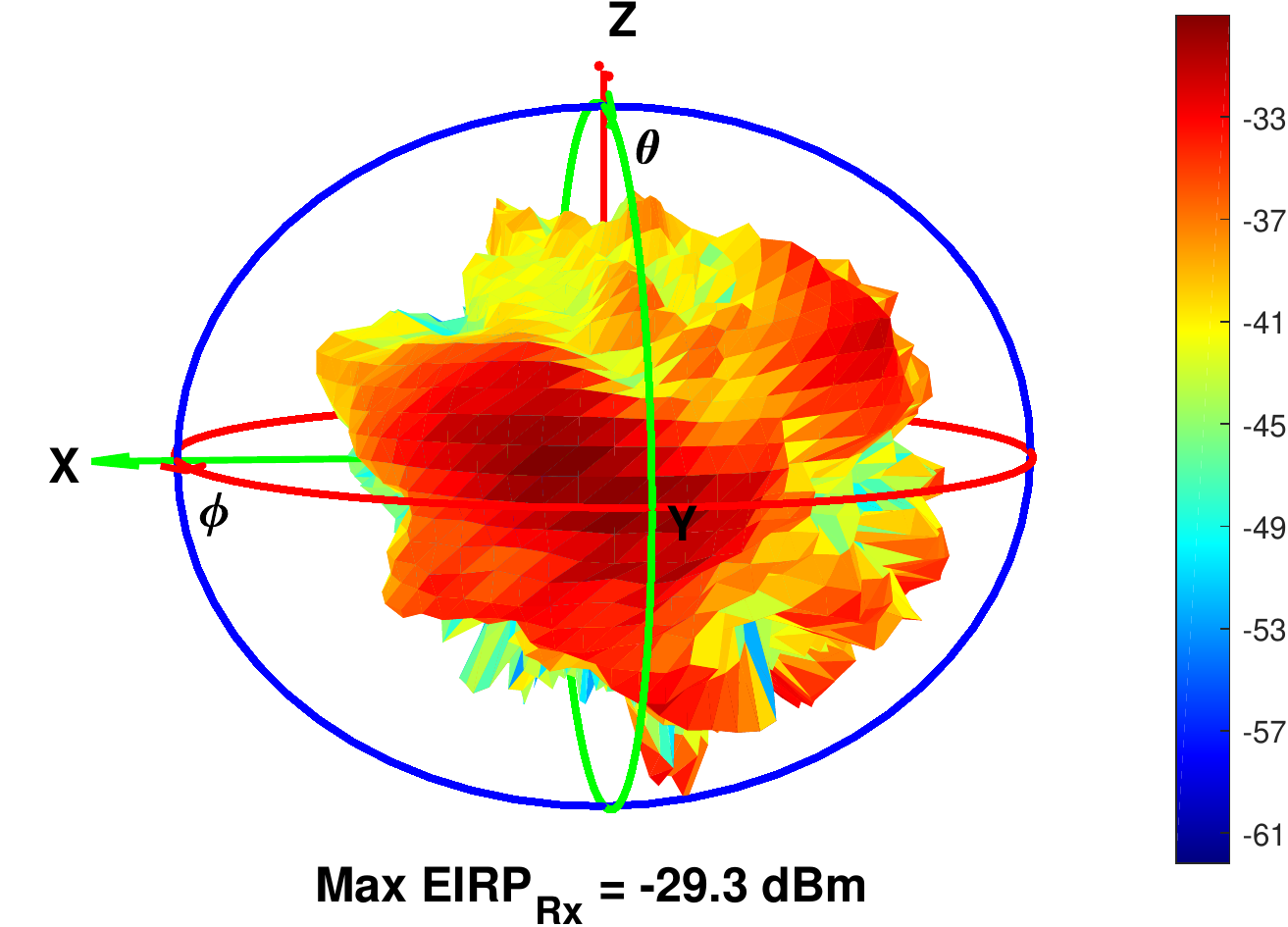}
&
\includegraphics[height=2.0in,width=2.2in]{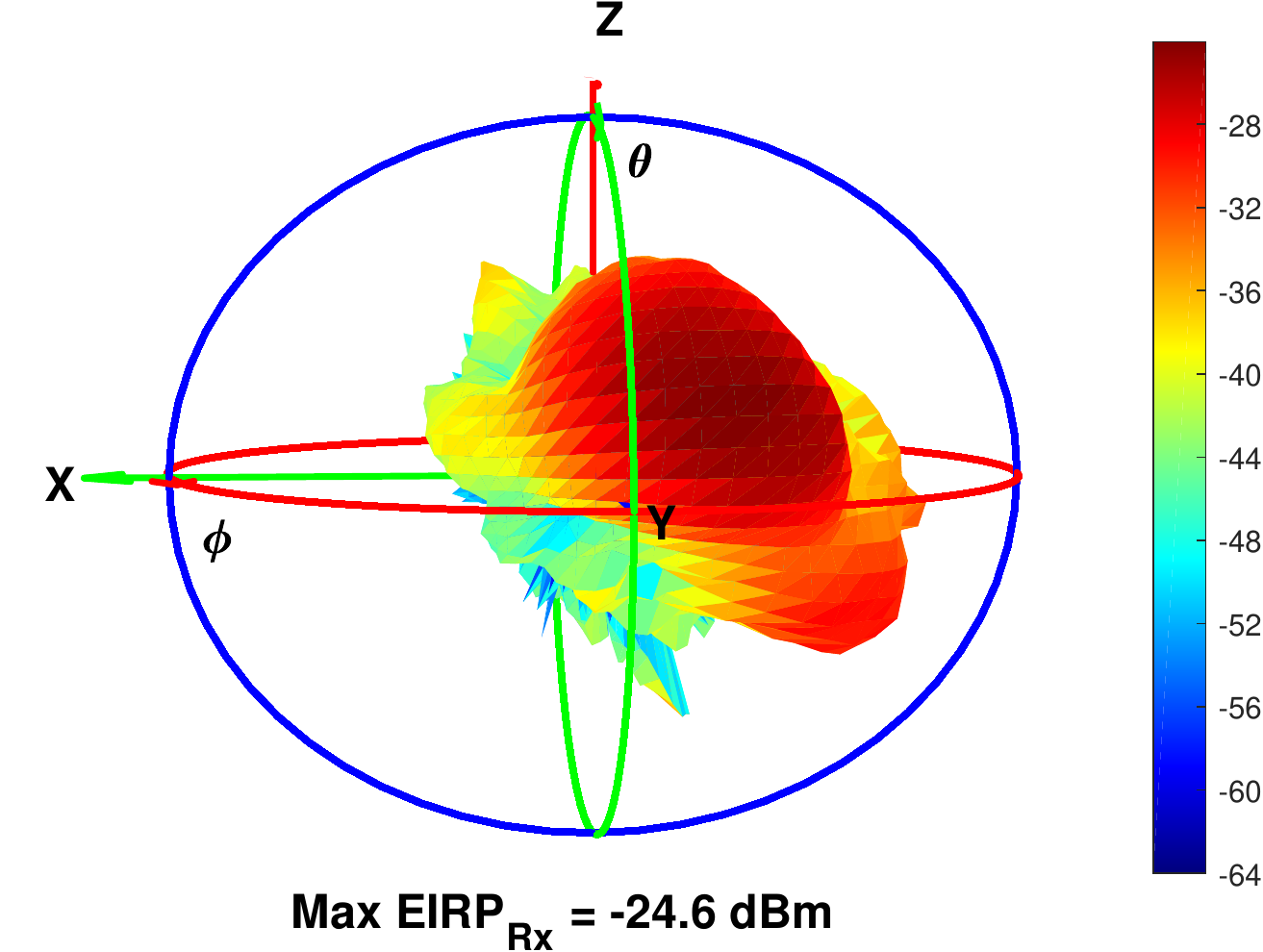}
\\
(d) & (e) & (f)
\\
\includegraphics[height=2.0in,width=2.2in]{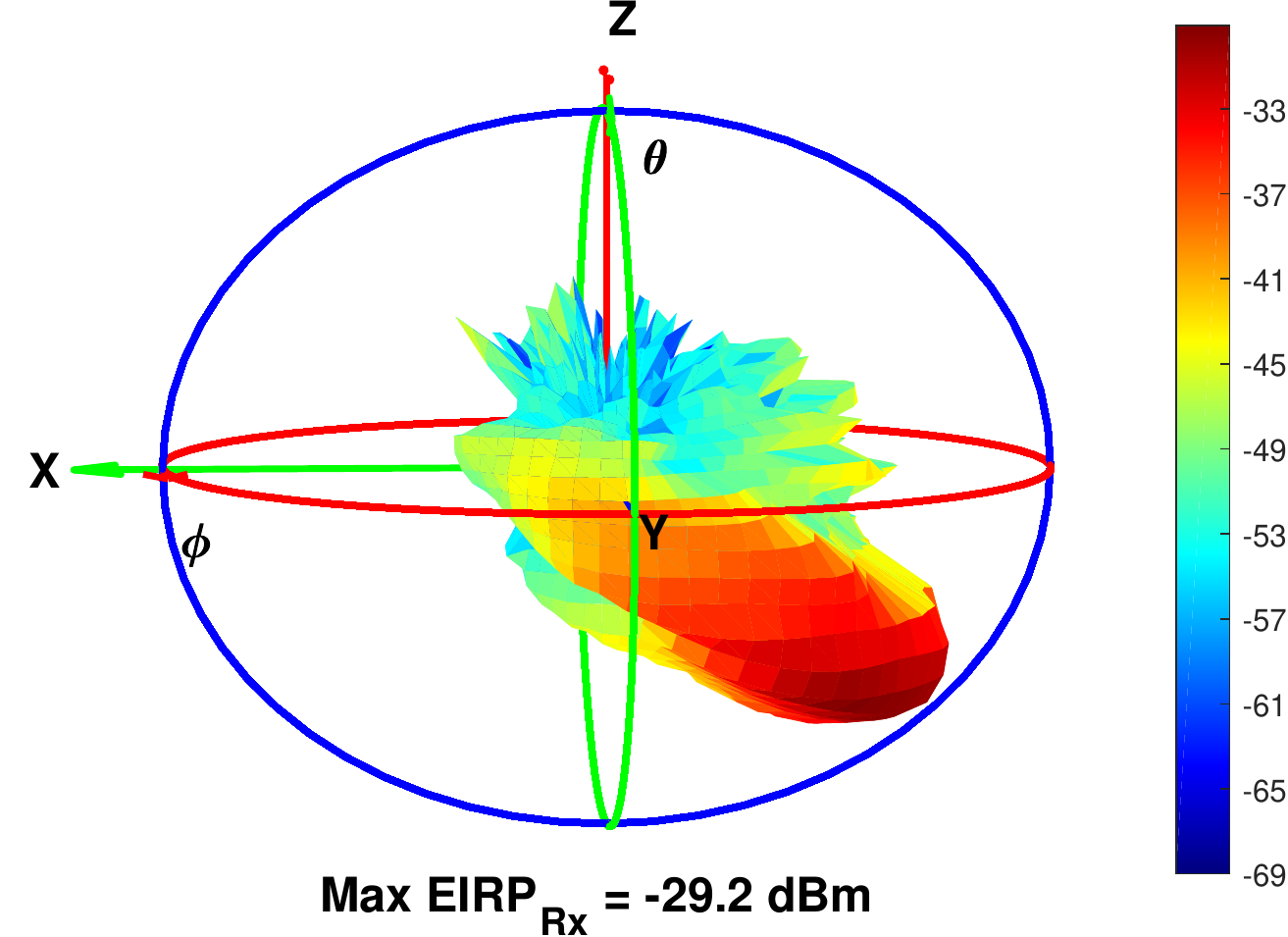}
&
\includegraphics[height=2.0in,width=2.2in]{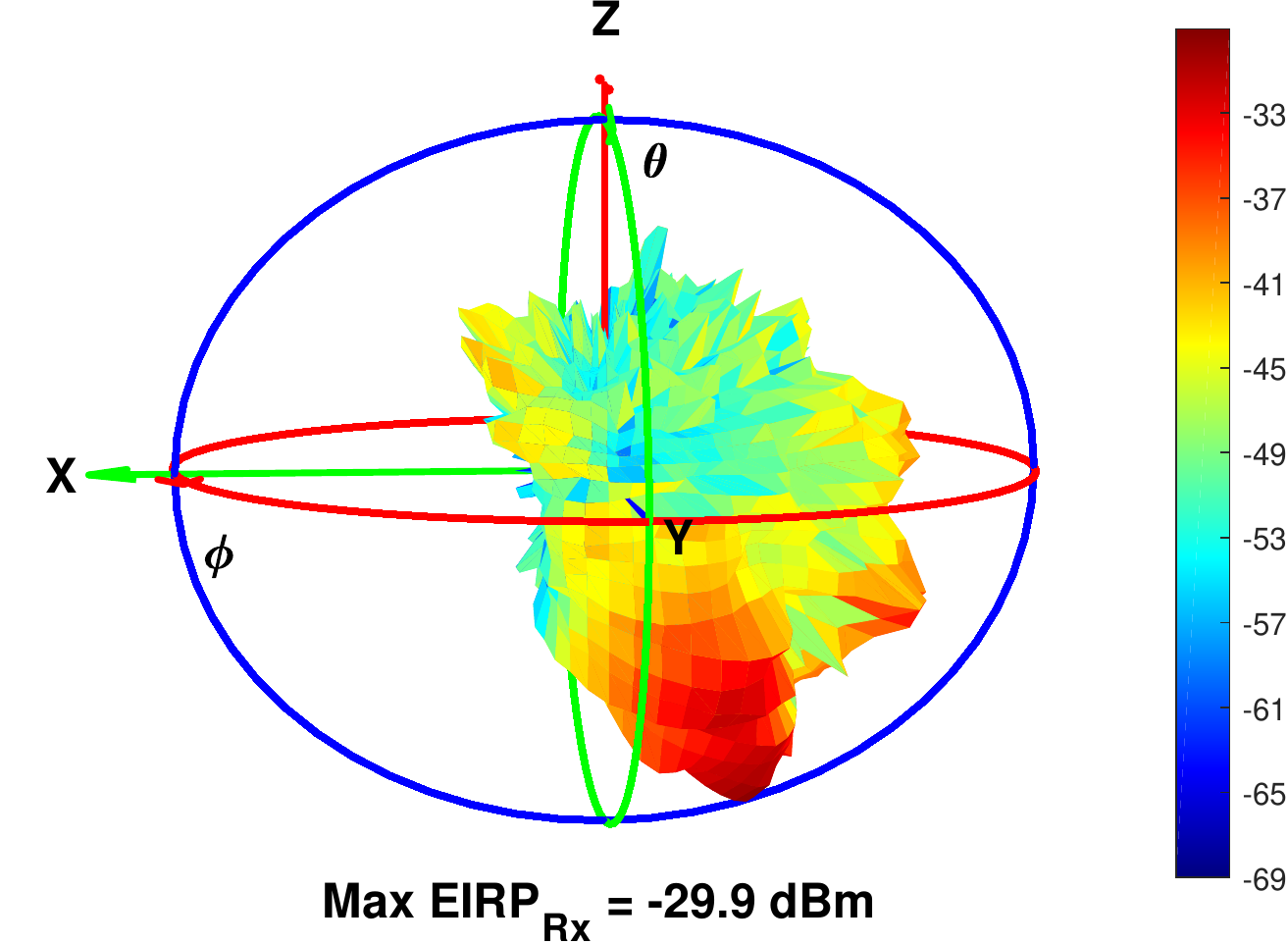}
&
\includegraphics[height=2.0in,width=2.2in]{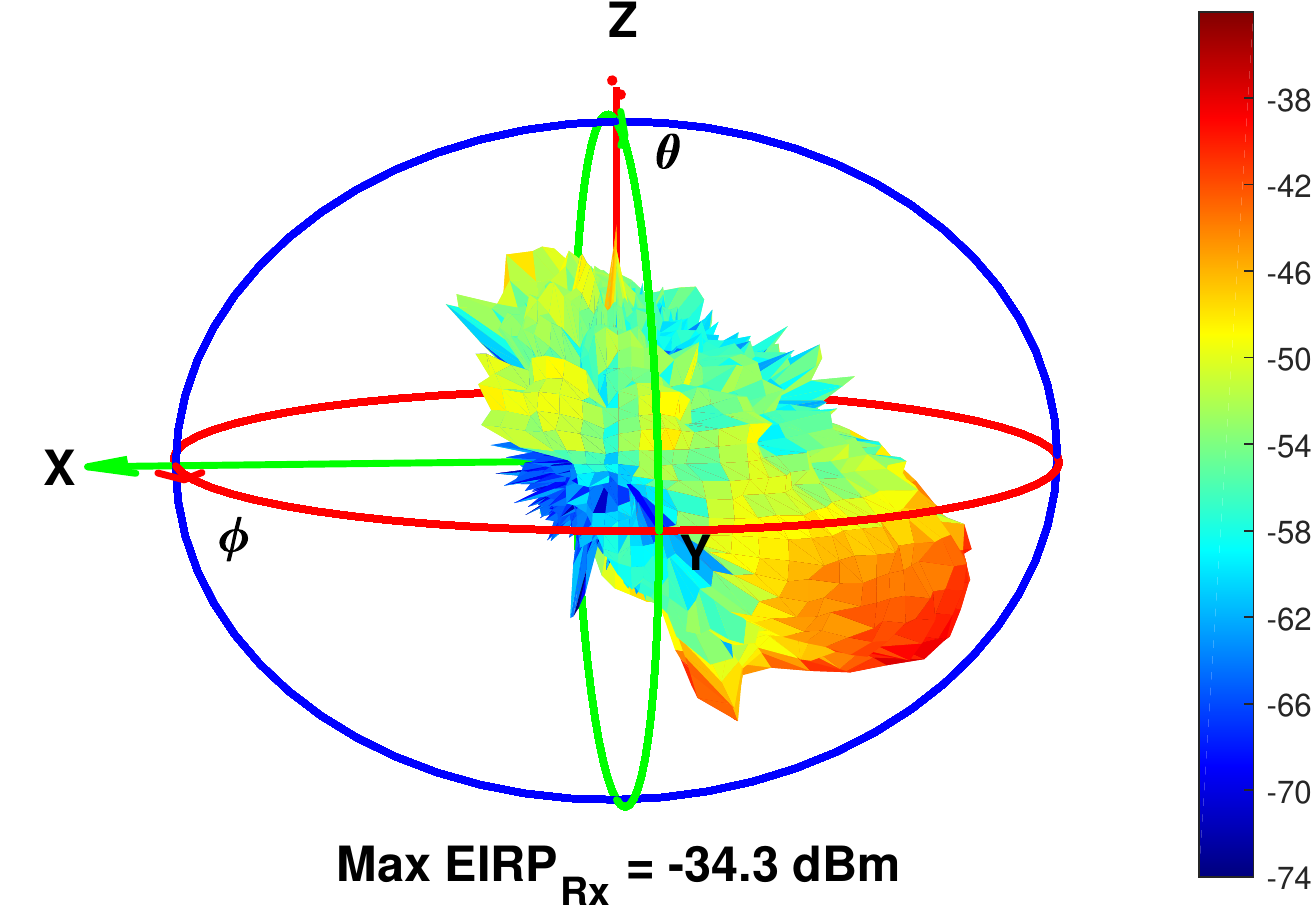}
\\
(g) & (h) & (i)
\end{tabular}
\caption{\label{fig_beampatterns_4by1patch}
Beam patterns in (a)-(c) Freespace, (d)-(f) with a hand phantom and (g)-(i) with a
true hand holding for the three beams considered for the $4 \times 1$ patch subarray.
}
\end{center}
%\vspace{-5mm}
\end{figure*}

\begin{table*}[htb!]
\caption{EIRP-Based Comparisons for the Hard Hand Grip (Study 1)}
\label{table_blockage}
\begin{center}
\begin{tabular}{|l|c| c| c|c|c| c|c|c|| c|c|c||}
\hline
\multirow{5}[6]{0.2cm}{\rotatebox[origin=c]{90}{Study 1}} &
& \multicolumn{7}{c||}{Spherical Coverage Lost}
& \multicolumn{3}{c||}{Loss (in dB)} \\ \cline{2-12}
& EIRP & Freespace
& \multicolumn{3}{c|}{Hand Phantom (in $\%$)}
& \multicolumn{3}{c||}{True Hand %Holding
(in $\%$)}
& Percentile & Hand Phantom & True Hand %Holding
\\ \cline{4-12}
& (in dBm) & (in $\%$) & & Abs. & Rel. & & Abs. & Rel. & $90$ & $3.1$ & $8.6$
\\ \cline{2-12}
& $>-35$ & $23.3$ &
$13.5$ & $9.7$ & $41.8$ &
$3.4$ & $19.8$ & $85.3$ &
$80$ & $3.4$ & $10.6$ \\ \cline{2-12}
%%%
& $>-40$ & $41.8$ &
$28.9$ & $12.9$ & $31.0$ &
$12.1$ & $29.7$ & $71.0$ &
$50$ & $3.4$ & $12.2$ \\ \cline{2-12}
%%%
& $>-45$ & $65.0$ &
$48.5$ & $16.6$ & $25.5$ &
$21.7$ & $43.4$ & $66.7$ &
$20$ & $2.0$ & $17.2$ \\
\hline \hline
\end{tabular}
%\end{center}
%{\vspace{0.1in}}
%%%%%%%%%%%%%%%%
%\begin{center}
\end{center}
\end{table*}
%\end{center}

\subsection{Beamforming Performance}
\label{sec3a}

%\subsection{$4 \times 1$ Patch Subarray with a Hard Hand Grip}
%\noindent {\bf \em \underline{Beamforming Codebook:}}
To understand the implications of blockage, we consider a codebook of three beams for the
$4 \times 1$ subarray. We consider a small codebook size of three since practical UE codebook
constraints are determined by low latency requirements for initial link
acquisition~\cite{raghavan_jstsp}. In the case of the $4 \times 1$ patch subarray, these
three beams are chosen to steer energy along the boresight of the array, $+30^{\sf o}$ to
the boresight and $-30^{\sf o}$ to the boresight, respectively. The beam patterns in
Freespace (over a sphere around the UE) are illustrated for these three beams in
Figs.~\ref{fig_beampatterns_4by1patch}(a)-(c), respectively. Note that in addition to
the correct orientation of the beam patterns (relative to the coordinate system in
Fig.~\ref{fig_hand_phantoms}(b)), the beam pattern of each beam is reasonably {\em regular}
and conforming to its theoretically expected performance~\cite{balanis}. Further, the
beamwidth of each beam is $\approx 25^{\sf o}$-$30^{\sf o}$ suggesting that these three
beams can cover a $75^{\sf o}$-$90^{\sf o}$ spatial area in one dimension, which is
typically the coverage area of a linear array at millimeter wave carrier frequencies.

\begin{figure*}[htb!]
\begin{center}
\begin{tabular}{cc}
\includegraphics[height=1.9in,width=2.8in]{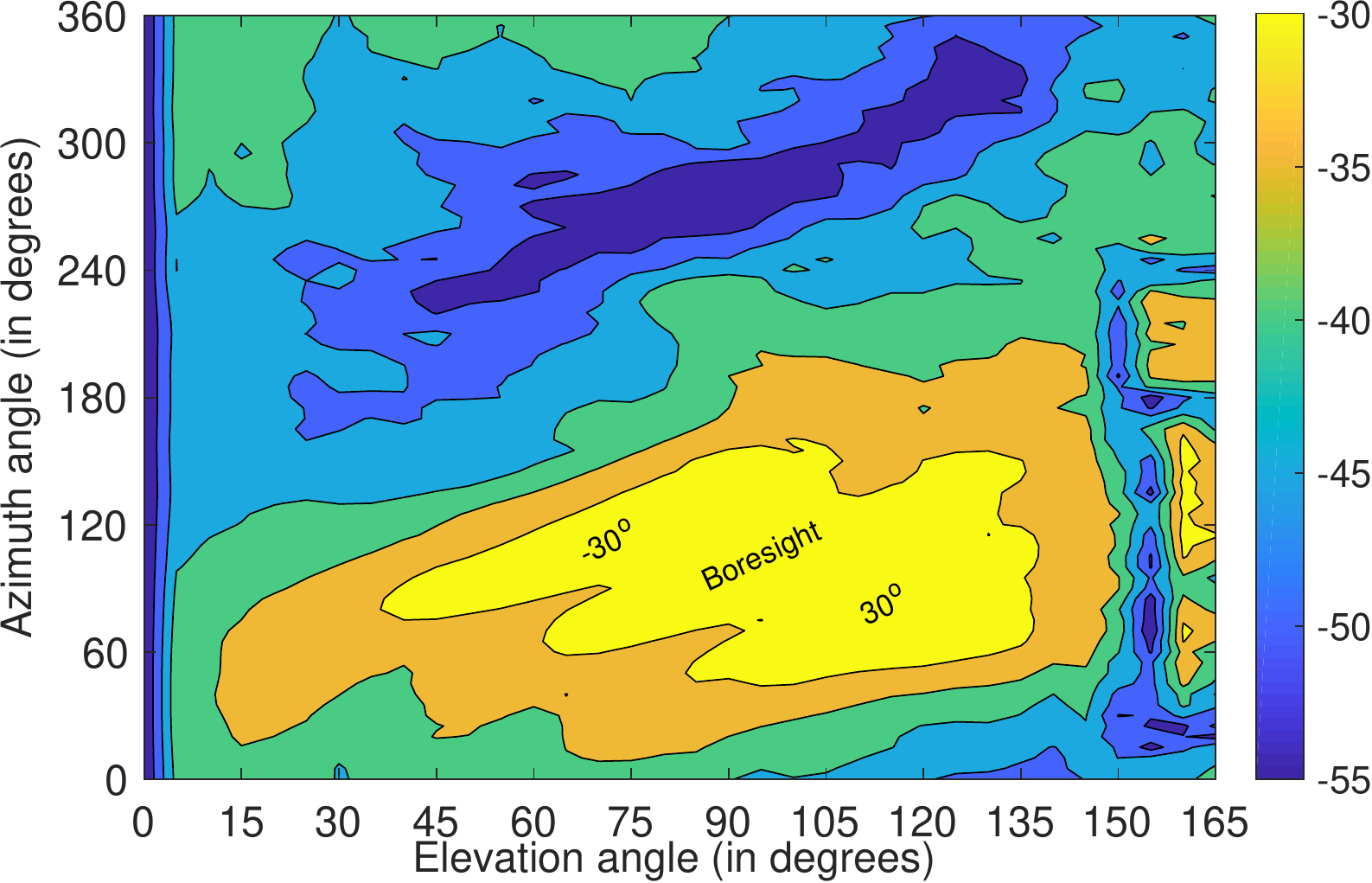}
&
\includegraphics[height=1.9in,width=2.8in]{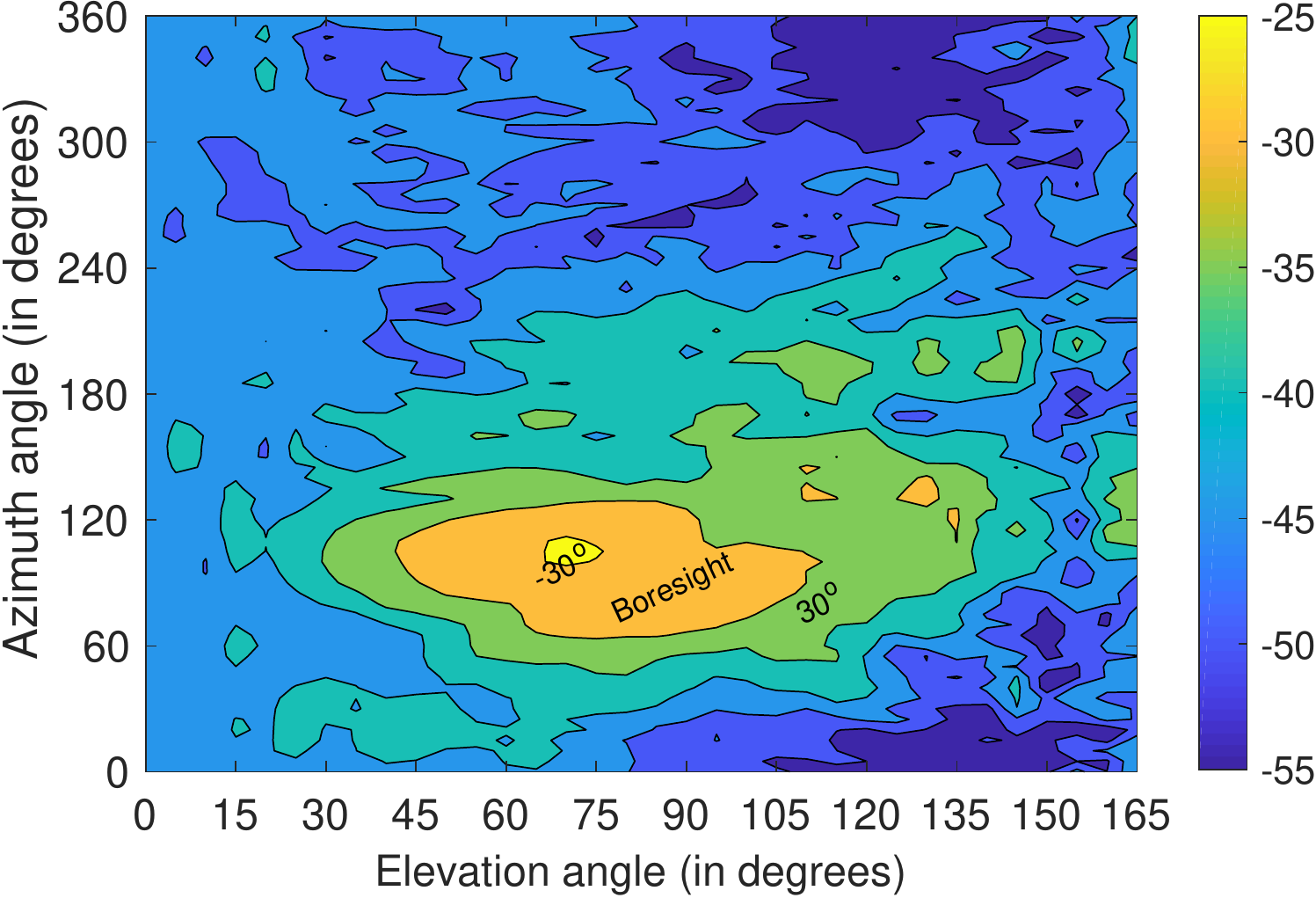}
\\
(a) & (b) \\
\includegraphics[height=1.9in,width=2.8in]{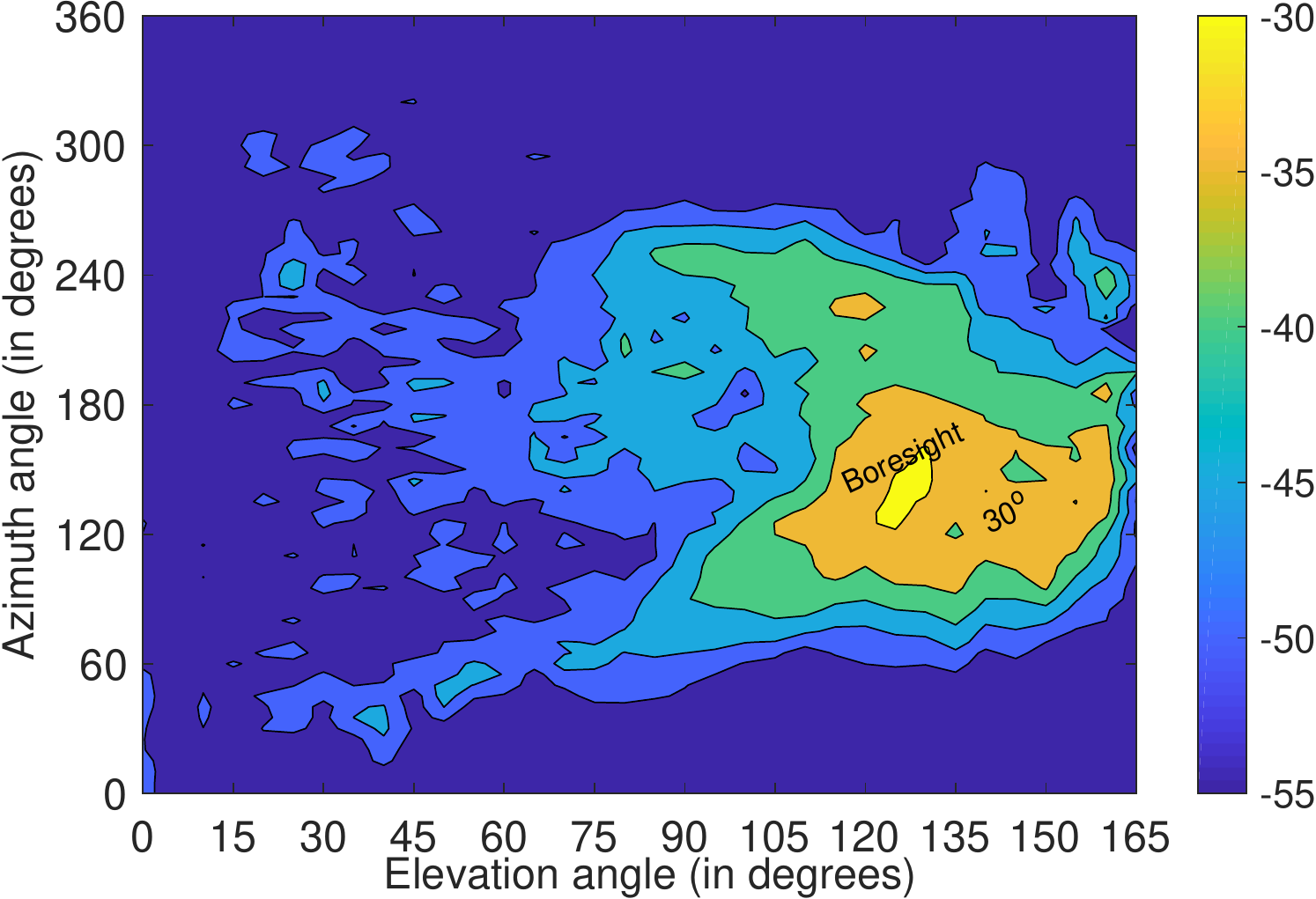}
&
\includegraphics[height=1.9in,width=2.8in]{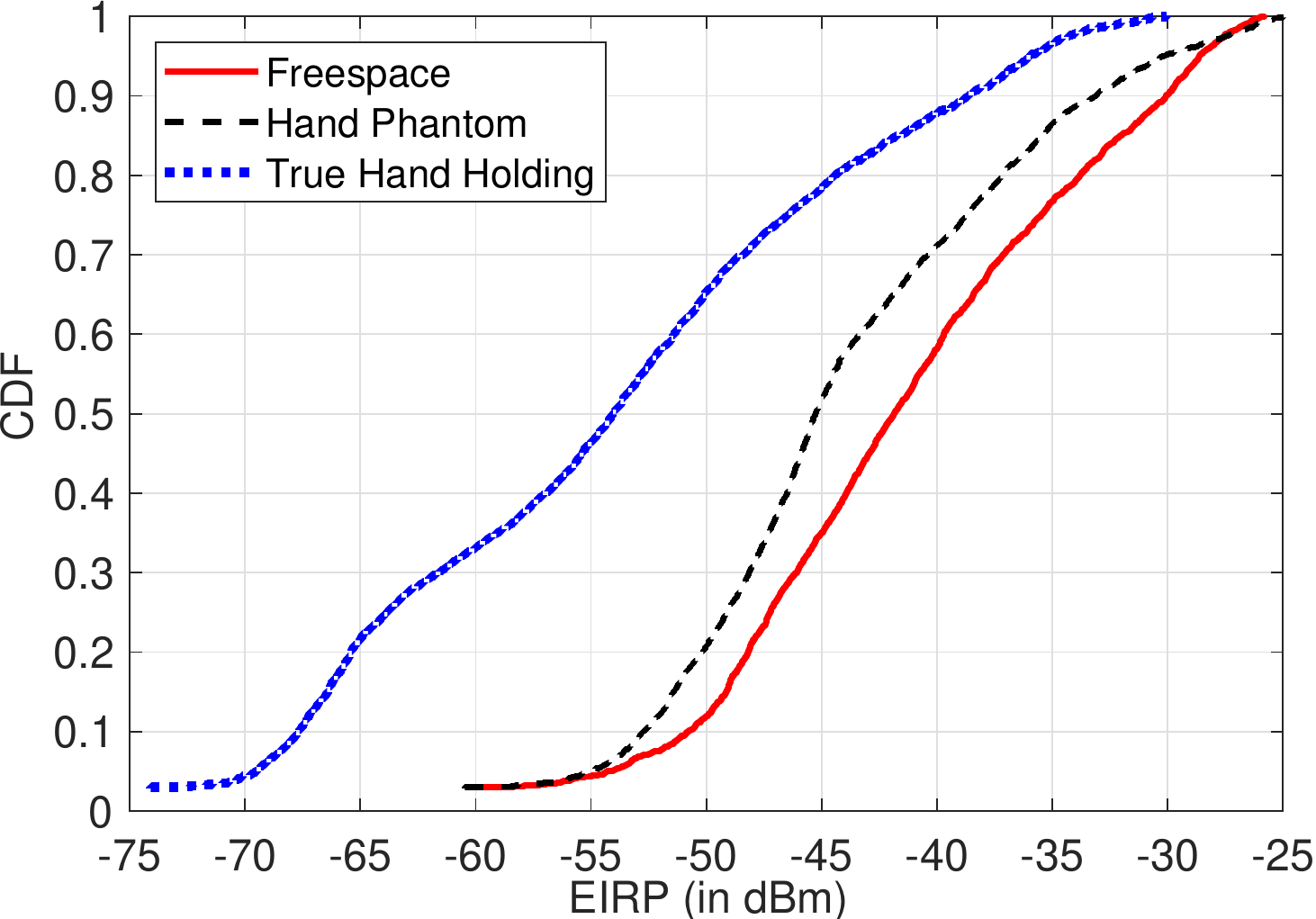}
\\
(c) & (d)
\end{tabular}
\caption{\label{fig_overlay_4by1patch}
(a)-(c) Overlay plots of the best of the three beam patterns in Freespace, with hand phantom and true hand
holding for the $4 \times 1$ patch subarray.
(d) CDFs of EIRP for different modes in the $4 \times 1$ patch subarray.
}
\end{center}
%\vspace{-5mm}
\end{figure*}

In a beamformed realization, an overlay plot of the beam pattern over the sphere due to the
best of the three beams is important. Such a characterization, by way of comparison without
and with blockage, also allows us to understand the impact of blockage on beamformed performance.
Fig.~\ref{fig_overlay_4by1patch}(a) illustrates this overlay plot for the $4 \times 1$
patch subarray in Freespace. In this plot, the behavior over the sphere is plotted as a
two-dimensional plane (over $\phi$-$\theta$ where $\phi$ and $\theta$ are the azimuth
and elevation angles, respectively). %and the $2 \times 1$ dipole subarray over Freespace, respectively. From both these
From this plot, we observe that this subarray is well-designed to ensure good coverage over at
least a $90^{\sf o} \times 60^{\sf o}$ coverage region in Freespace, which is typical for
antennas at millimeter wave carrier frequencies.

%\subsection{Beam Patterns with Hand Phantom and Real Hand Holding}
We next consider the behavior of the beam weights used with the patch subarray
with the hand phantom and a true hand holding the UE. Figs.~\ref{fig_beampatterns_4by1patch}(d)-(f)
and (g)-(i) plot the beam patterns of the three beams in these two setups, respectively. The
overlay plots of the best of the three beams' beam patterns are plotted in these two scenarios
in Figs.~\ref{fig_overlay_4by1patch}(b) and~(c), respectively. Note
that Fig.~\ref{fig_overlay_4by1patch}(b) shows that some beams have
better performance with the hand phantom than in Freespace which is possible due to
reflections by the hand phantom, whereas Fig.~\ref{fig_overlay_4by1patch}(c)
shows that the impact of the beam which is at $-30^{\sf o}$ off boresight is rarely observed.

From these plots, we also observe that while the hand phantom {\em irregularly} distorts the
beam patterns in Freespace, this distortion is still rather minor. In particular, the peak EIRP is
distorted by $0.5$ dB, $2.3$ dB and $1.1$ dB (gains in some cases) for the three beams,
respectively. On the other hand, the distortion with a true hand holding is significant (peak
EIRP distortion of $3.4$ dB, $2.9$ dB and $8.6$ dB) with considerable signal deterioration
observed over the beam patterns in Freespace. Specifically, the beam which is $+30^{\sf o}$ off
boresight steers energy towards the torso, hips and stomach of the human holding the UE and is
thus less impacted in terms of signal distortion relative to the beam which is $-30^{\sf o}$ off
boresight, which steers energy towards the face and shoulder and is thus significantly impacted
by the human. The wide variations between the hand phantom behavior and a true human holding the
UE indicates that the use of the hand phantom to replicate\footnote{Blockage studies with
non-human intervention in the lab/chamber at millimeter wave frequencies would be of great
interest if the hand phantom results faithfully replicate true hand holding results. The studies
in this paper show that there is a considerable gap between these two sets of studies, which
justifies the need for further work in careful tuning of the hand phantom to match the results
from true hand holding studies. At the very least, more studies are necessary to
understand if/when hand phantoms can replicate true hand holding results.} true hand holding
studies is {\em risky} in terms of drawing
meaningful lessons. This is because the hand phantom can at best replicate the effects of the
hand with careful tuning, but not capture the effects of the body. Thus, true hand/body-based
blockage studies are necessary to understand the implications of blockage.

\subsection{Estimating the Loss Region}
\label{sec3b}
To understand the implications of beamforming, we now plot the CDFs of the EIRP as seen
over the sphere (weighted by $\sin(\theta)$) in Freespace and with the hand (both hand
phantom and the true hand holding). The weighting by $\sin(\theta)$ is essential since
the sample points in the $\phi$-$\theta$ plane are uniform (at steps of $5^{\sf o}$)
leading to a crowding of points near the poles, which needs to be adjusted by the Jacobian
of the coordinate transformation from a Cartesian/rectangular coordinate system to a
spherical coordinate system~\cite{vasanth_tcom2019}. This plot in Fig.~\ref{fig_overlay_4by1patch}(d)
shows that there are two ways to interpret the CDF data. In the first view, the true hand
holding leads to a $20\%$-$45\%$ (absolute) spherical coverage loss at different EIRP
levels. In the second view, the true hand leads to a signal strength degradation of
$8.5$-$17$ dB at different percentile points. Reinforcing the prior observations on the
mismatch of the hand phantom in capturing blockage performance, we note that the hand
phantom leads to a $9.5\%$-$17\%$ (absolute) spherical coverage loss at different EIRP
levels, or an equivalent $2$-$3.5$ dB loss at different percentile points.
Table~\ref{table_blockage} provides a summary of the absolute and relative spherical coverage
losses at different EIRP levels as well as the loss seen at different percentile points.

Given such a wide range of losses at different percentile points, it is reasonable to
ask as to what is a good model for spherical coverage loss and/or blockage loss.
%Section~\ref{sec4} deals with this question in more detail.
We now deal with this question in more detail.
To understand the impact of blockage, two broad questions are laid out in prior works:
\begin{itemize}
\item What is the {\em Region-of-Interest} (RoI) in terms of blockage's impact?
\item What is the loss seen over this RoI? Can this loss be modeled as an appropriate
stochastic distribution?
\end{itemize}
In the sequel, we show that the above view is quite {\em simplistic} in terms of characterizing
blockage performance. Specifically, we show that blockage does not just lead to losses
over the RoI, but can also lead to gains due to reflection of signals from the fingers,
palm and the hand. The precise nature and scale of the reflection gains depends on the
type of hand grip and orientation, user-specific skin properties, etc. Thus, to truly
understand the implications of blockage in terms of physical layer performance, we need
to define the RoI carefully. Towards this goal, we define multiple such RoIs and show the
broad utility of two specific RoI definitions.

\begin{figure*}[htb!]
\begin{center}
\begin{tabular}{cc}
\includegraphics[height=1.9in,width=2.8in]{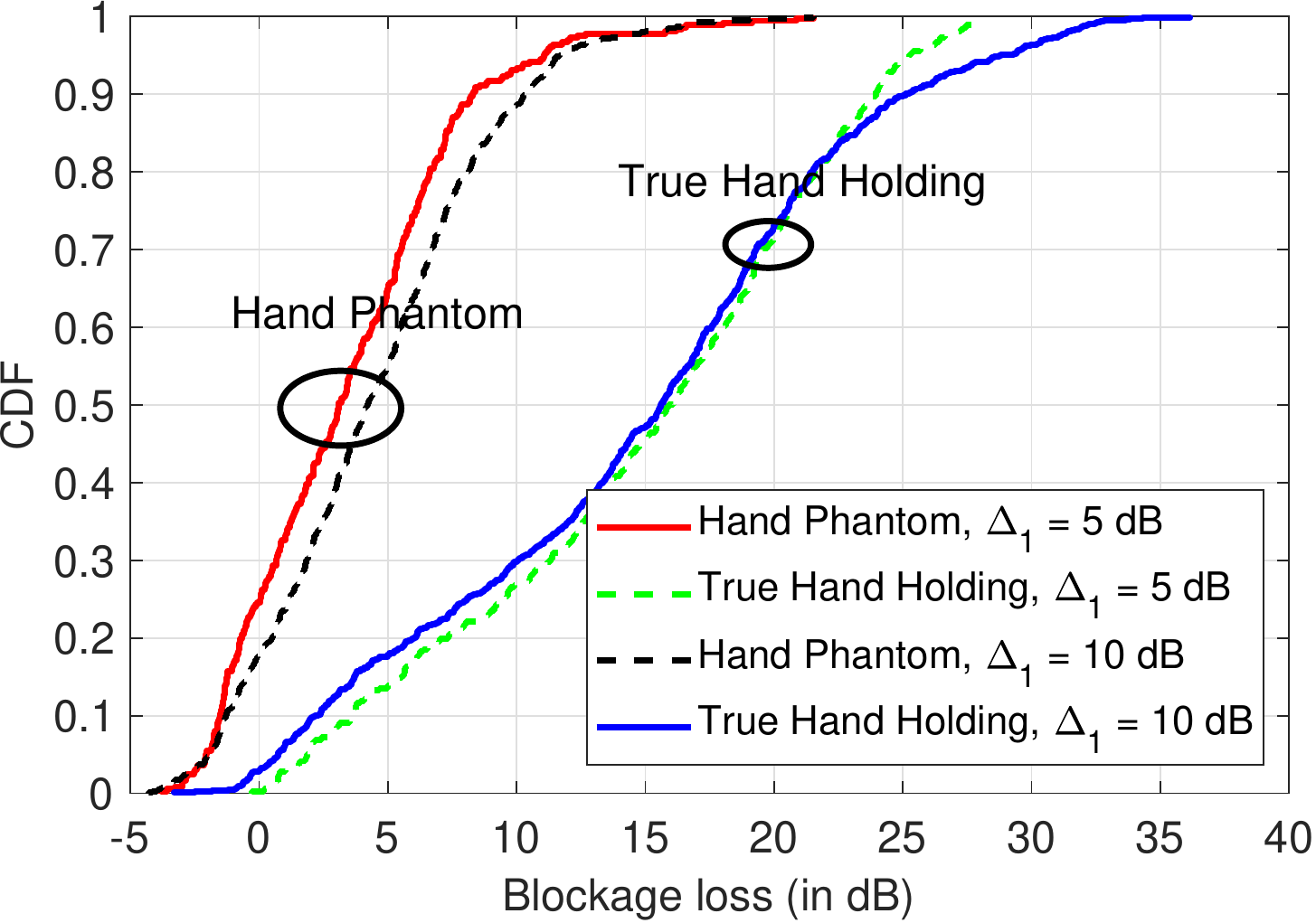}
&
\includegraphics[height=1.9in,width=2.8in]{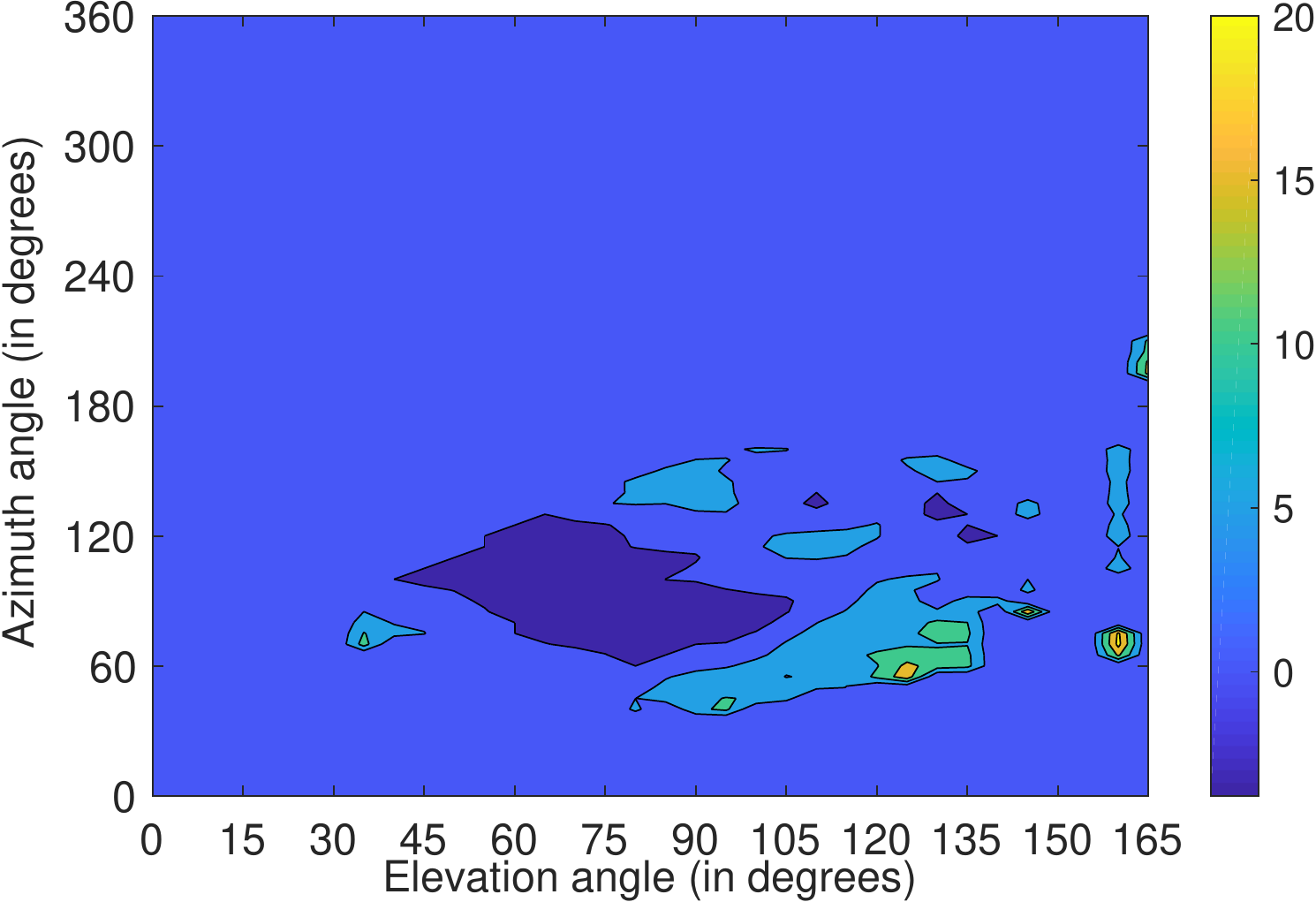}
\\
(a) & (b)
\\
\includegraphics[height=1.9in,width=2.8in]{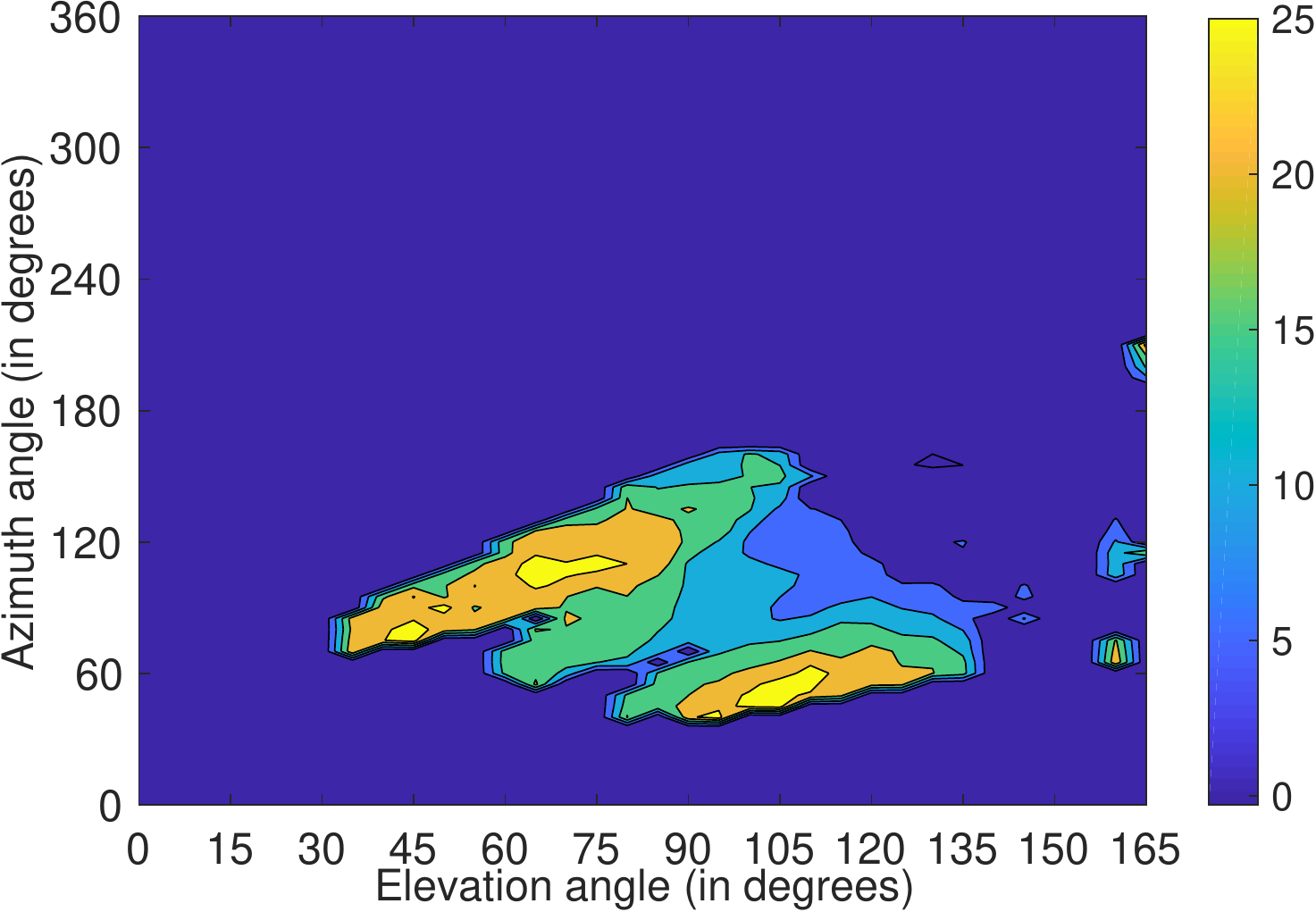}
&
\includegraphics[height=1.9in,width=2.8in]{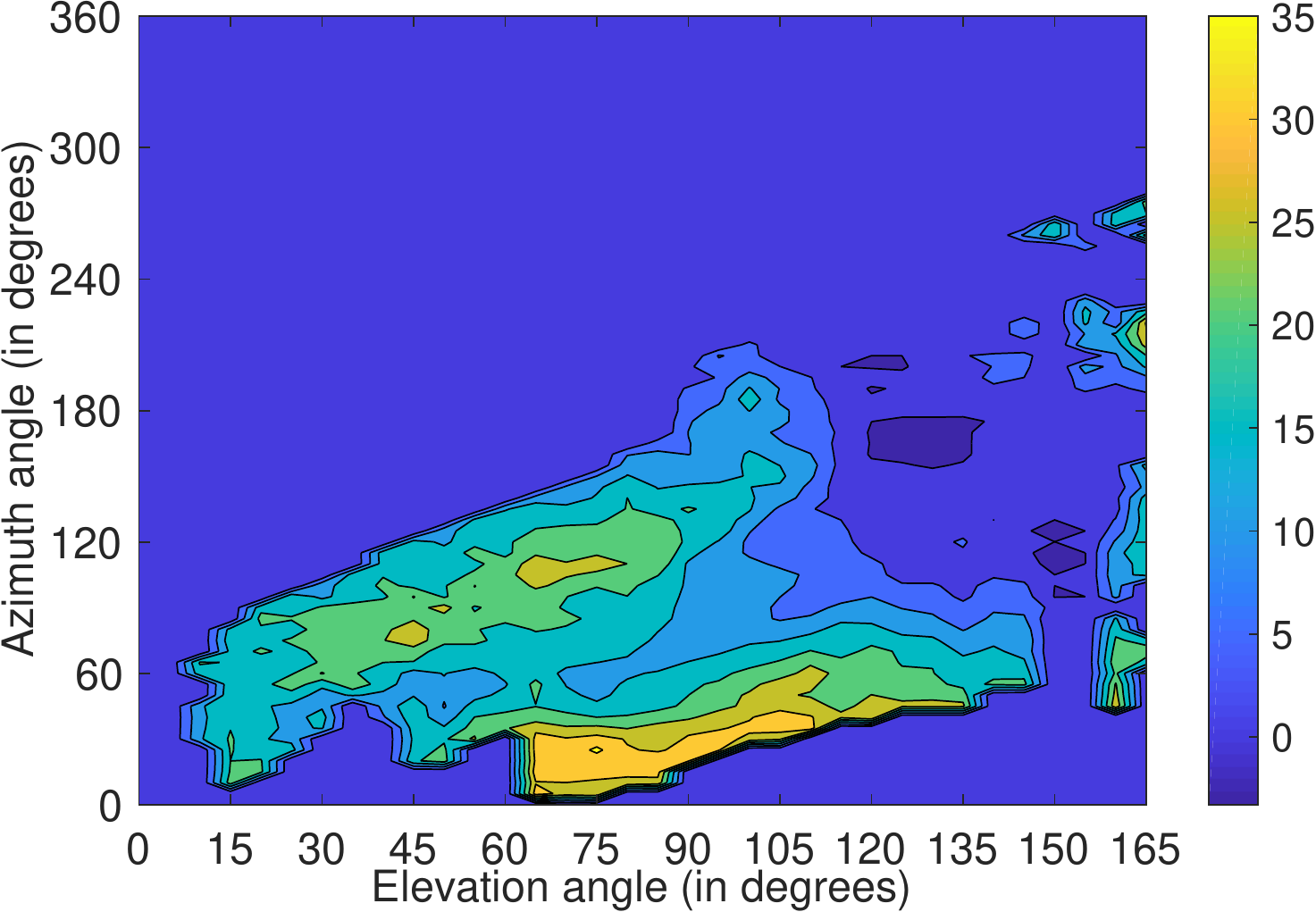}
\\
(c) & (d)
\end{tabular}
\caption{\label{fig_CDFs_RoI} (a) CDFs of blockage loss with the hard hand grip over
${\cal R}_1$ with $\Delta_1 = 5$ and $10$ dB. Blockage behavior over the sphere with
(b) hand phantom and (c) true hand holding for ${\cal R}_1$ with $\Delta_1 = 5$ dB.
(d) Blockage behavior over the sphere with true hand holding
for ${\cal R}_1$ with $\Delta_1 = 10$ dB. }
\end{center}
%\vspace{-5mm}
\end{figure*}

\begin{figure*}[htb!]
\begin{center}
\begin{tabular}{cc}
\includegraphics[height=1.9in,width=2.8in]{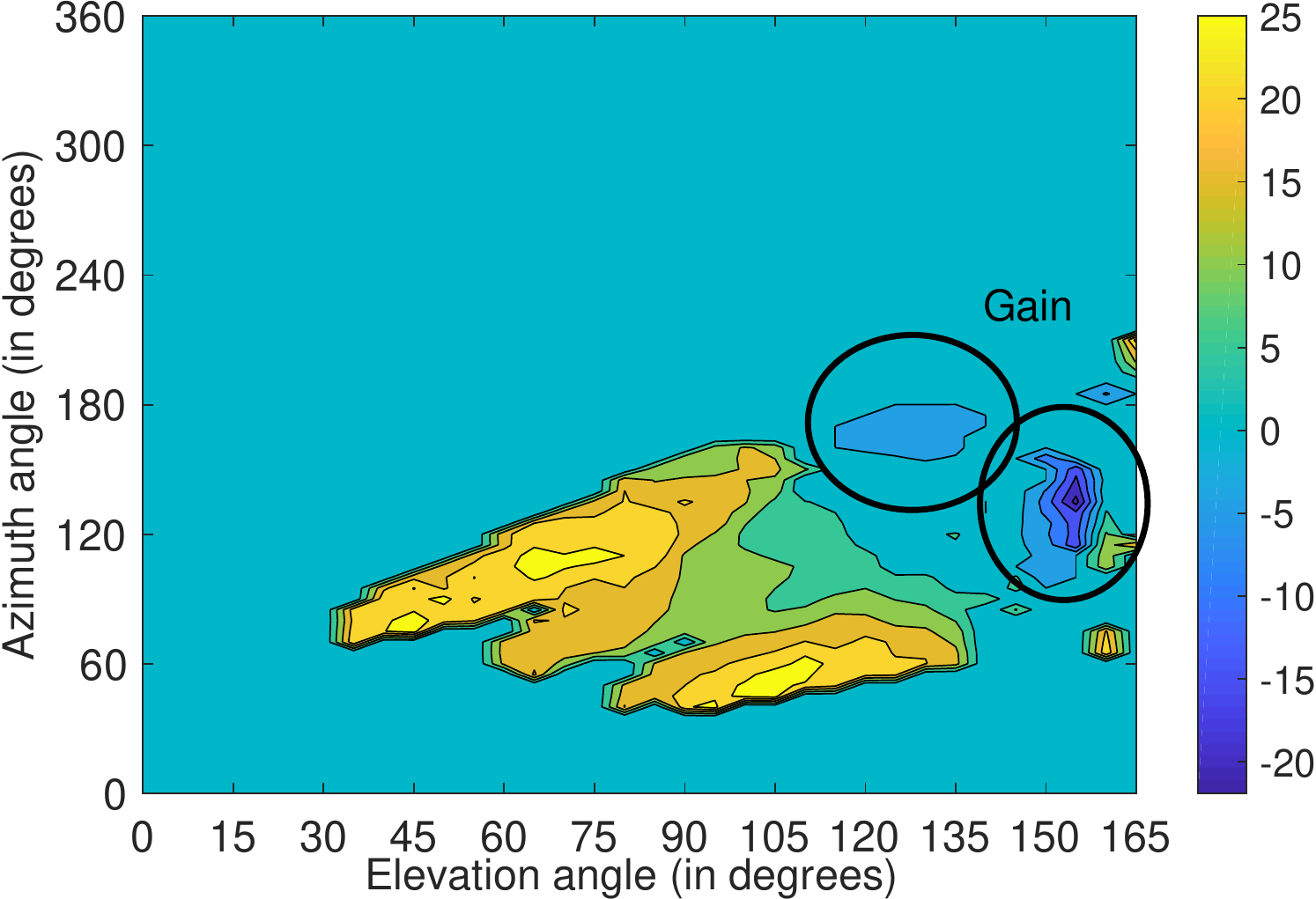}
&
\includegraphics[height=1.9in,width=2.8in]{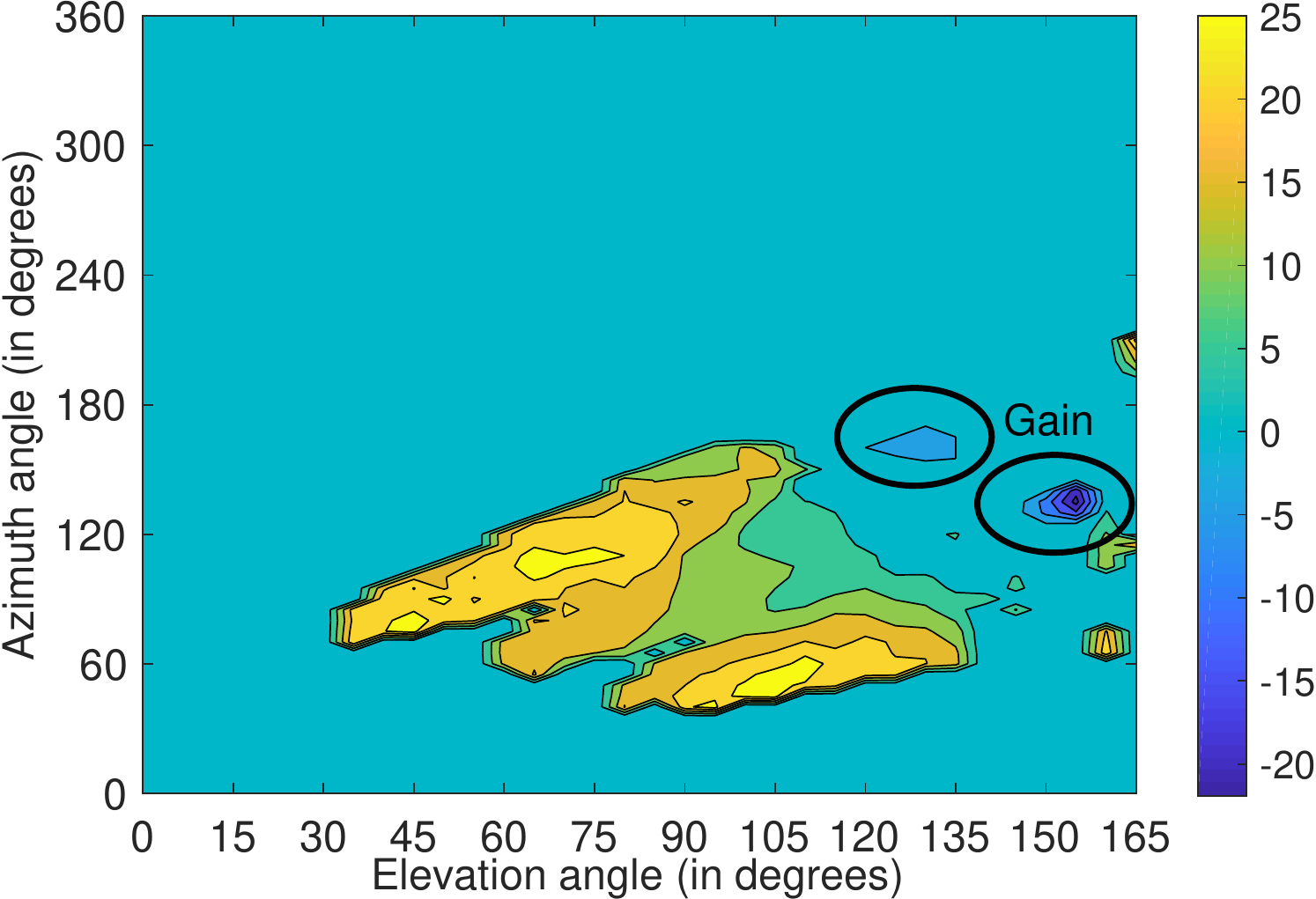}
\\
(a) & (b)
\\
\includegraphics[height=1.9in,width=2.8in]{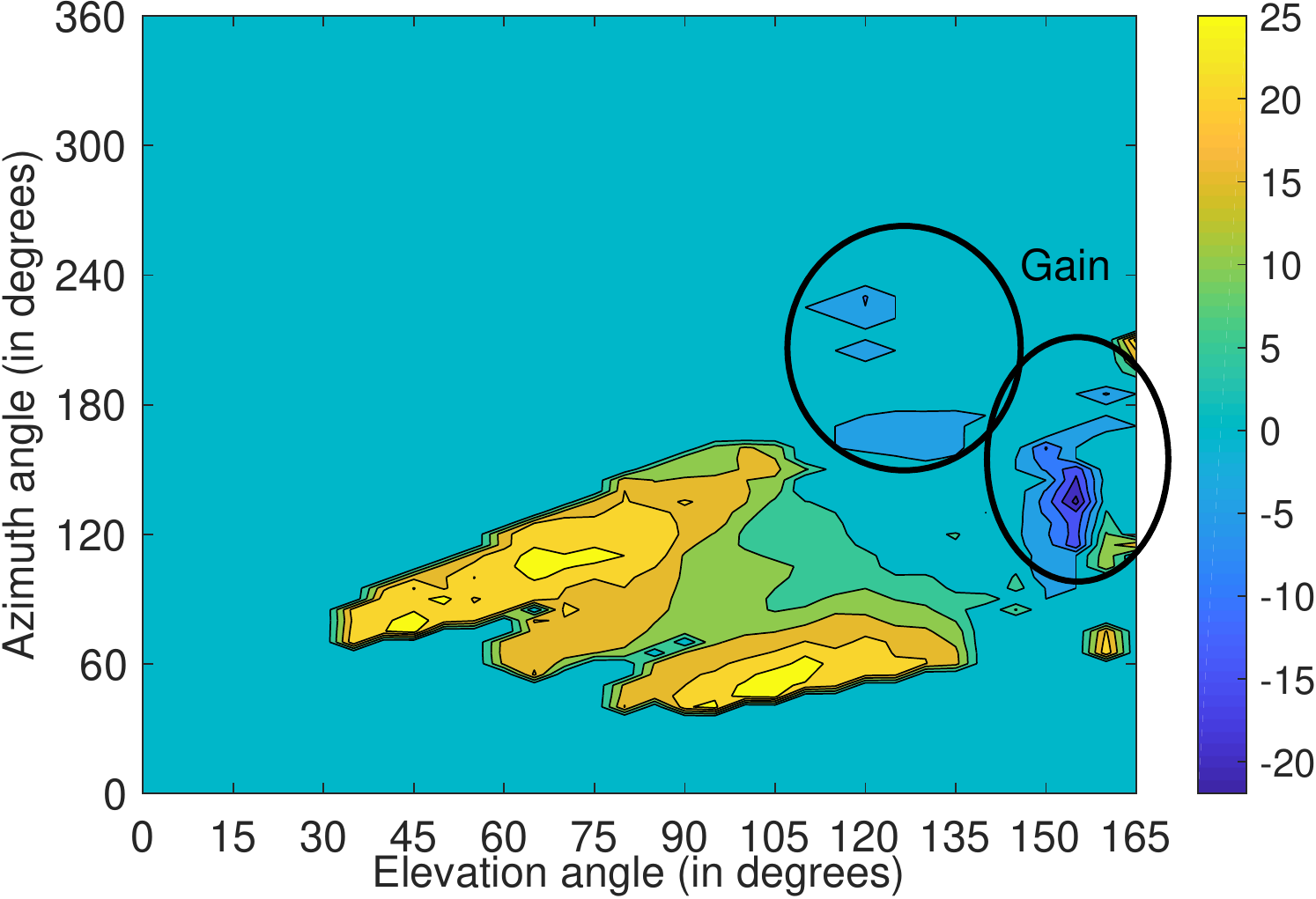}
&
\includegraphics[height=1.9in,width=2.8in]{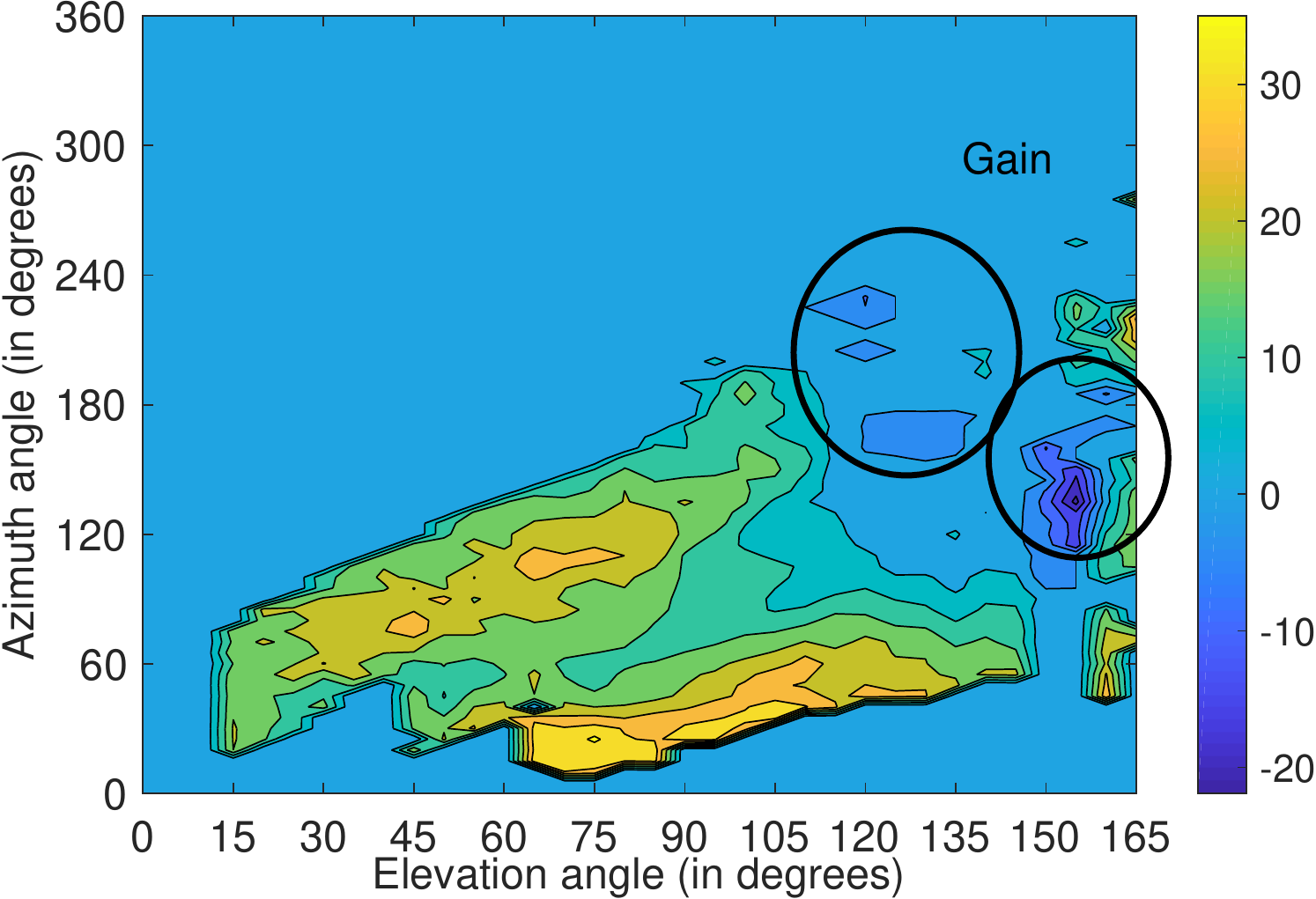}
\\
(c) & (d)
\end{tabular}
\caption{\label{fig_CDFs_RoI_others}
CDFs of blockage loss with the hard hand grip over (a) ${\cal R}_2$ with $\Delta_1 = \Delta_2 = 5$
dB, (b) ${\cal R}_3$ with $\Delta_1 = \Delta_3 = 5$ dB,
(c) ${\cal R}_4$ with $\Delta_1 = 5$ dB and $\Delta_4 = -35$ dBm, and
(d) ${\cal R}_5$ with $\Delta_5 = -35$ dBm.}
\end{center}
%\vspace{-5mm}
\end{figure*}

Let $G(\theta, \phi)$ denote the beamforming array gain (in dB) seen in Freespace in a
certain direction $(\theta, \phi)$. Let $G_{\sf max}$ (also in dB) denote the maximum gain
in Freespace over all directions. That is,
\begin{eqnarray}
G_{\sf max} = \max_{\theta, \phi} G(\theta, \phi).
\nonumber
\end{eqnarray}
The typical definition of a RoI for a certain choice of threshold $\Delta_1$ (in dB)
is:
\begin{eqnarray}
{\cal R}_1(\Delta_1) = \left\{ (\theta, \phi) : G(\theta, \phi) \geq G_{\sf max} -
\Delta_1 \right\}.
\nonumber
\end{eqnarray}
That is, ${\cal R}_1(\Delta_1)$ captures the region where the Freespace gains are within
a fixed cutoff ($\Delta_1$ dB) of $G_{\sf max}$. It is important to note that this
definition of ${\cal R}_1$ only relies on the Freespace gain, and not on what happens
with hand/body blockage (which is quite na{\"i}ve from understanding the implications of
blockage). Further, this region does not have to be a rectangular/regular region in
$(\theta, \phi)$ nor does it have to be a single connected region. In general,
${\cal R}_1(\Delta_1)$ could be a union of multiple irregular regions.

We now consider multiple ways in which blockage performance can be incorporated into
the definition of a RoI by enhancing ${\cal R}_1(\Delta_1)$.
Let $G_{\sf body}(\theta, \phi)$ and $G_{\sf max, body}$ denote the gain seen with
hand/body holding the UE in a direction $(\theta, \phi)$ and the maximum of this gain
over all $(\theta, \phi)$. We can define the following four RoIs for different choices of
$\Delta_2, \Delta_3, \Delta_4$ and $\Delta_5$:
\begin{eqnarray}
{\cal R}_2(\Delta_1, \Delta_2) & = &
\left\{  (\theta, \phi) : G(\theta, \phi) \geq
G_{\sf max} - \Delta_1 {\hspace{0.05in}} {\sf or} {\hspace{0.05in}}
\right. \nonumber \\
& & \left. G_{\sf body}(\theta, \phi) \geq G_{\sf max, body} - \Delta_2
\right\} \nonumber \\
%%%%%
{\cal R}_3(\Delta_1, \Delta_3) & = &
\left\{  (\theta, \phi) : G(\theta, \phi) \geq
G_{\sf max} - \Delta_1 {\hspace{0.05in}} {\sf or} {\hspace{0.05in}}
\right. \nonumber \\
& & \left. G_{\sf body}(\theta, \phi) \geq G_{\sf max} - \Delta_3
\right\} \nonumber \\
%%%%%%%%%%%%%%
{\cal R}_4(\Delta_1, \Delta_4) & = &
\left\{  (\theta, \phi) : G(\theta, \phi) \geq
G_{\sf max} - \Delta_1 {\hspace{0.05in}} {\sf or} {\hspace{0.05in}}
\right. \nonumber \\
& & \left.
G_{\sf body}(\theta, \phi) \geq \Delta_4 \right\}
\nonumber \\
%%%%%%%%%%%%
{\cal R}_5(\Delta_5) & = &
\left\{  (\theta, \phi) : G(\theta, \phi) {\hspace{0.05in}} {\sf or}
{\hspace{0.05in}}
G_{\sf body}(\theta, \phi) \geq \Delta_5
\right\}.
\nonumber
\end{eqnarray}
The intuitive meaning of these RoIs is that in addition to the region captured by
${\cal R}_1$, they capture the following:
\begin{itemize}
\item
${\cal R}_2$ captures the additional region where the body reflects power.
\item
${\cal R}_3$ captures the additional region where the body reflects power as long as
that power is viable from a Freespace perspective.
\item
${\cal R}_4$ captures the additional region where the body reflects power that is
nominally good as described by $\Delta_4$ (where this $\Delta_4$ parameter is typically
chosen to meet some link budget constraint).
\item
${\cal R}_5$ captures the region where either Freespace- or hand/body blockage-based signals
are good as described by the {\em common} link margin threshold $\Delta_5$ (which is
chosen to meet some link budget constraint).
\end{itemize}
Note that for $\Delta_2 = \Delta_3$, ${\cal R}_3$ corresponds to a more aggressive
definition of RoI relative to ${\cal R}_2$ since $G_{\sf max} \geq G_{\sf body,max}$. On
the other hand, the definitions
of ${\cal R}_4$ and ${\cal R}_5$ strike a balance between the definitions of ${\cal R}_2$
and ${\cal R}_3$ with appropriate choices of $\Delta_4$ and $\Delta_5$. Also, note that
${\cal R}_4$ and ${\cal R}_5$ rely on link budget-driven parameters, while ${\cal R}_2$
and ${\cal R}_3$ do not. That said, by appropriately tuning $\Delta_2$, $\Delta_3$ and
$\Delta_4$ (in a non-link budget driven basis), ${\cal R}_2$, ${\cal R}_3$ and ${\cal R}_4$
can be made comparable with each other.

We now present the CDFs of the EIRP differential between the Freespace and %hand phantom/
true hand holding scenarios for the RoI corresponding to ${\cal R}_1$ with $\Delta_1 = 5$
dB and $\Delta_1 = 10$ dB in Fig.~\ref{fig_CDFs_RoI}(a). The RoI ${\cal R}_1$ with
$\Delta_1 = 5$ dB captures $\approx 14.6\%$ of the sphere with a mean, median and standard
deviation of the loss with true hand holding being $14.9$ dB, $16.0$ dB and $7.3$ dB,
respectively. Similarly, for ${\cal R}_1$ with $\Delta_1 = 10$ dB, the RoI captures $\approx 32.6\%$ of the
sphere with the mean, median and standard deviation of the loss with true hand holding being
$14.7$ dB, $15.6$ dB and $8.4$ dB, respectively. In both scenarios, there is a wide
discrepancy in terms of the CDF behavior for the true hand holding relative to the hand
phantom (which is not entirely surprising). Further, the EIRP differential over these
regions for the best of the three beams are plotted in the $\phi$-$\theta$ plane in
Figs.~\ref{fig_CDFs_RoI}(b)-(d), respectively. These plots correspond to the hand
phantom's behavior with $\Delta_1 = 5$ dB and true hand holding for $\Delta_1 = 5$ and
$10$ dB, respectively. Clearly, these plots show that the true hand holding behavior
seen over ${\cal R}_1$ is mostly loss, which is typical of blockage performance
characterization in prior works.

For other RoIs, Fig.~\ref{fig_CDFs_RoI_others} provides a representative plot of blockage
performance with some choices of parameters defining these RoIs. From these plots, we observe
that there are some regions of the sphere where hand reflections can lead to substantial
gains (regions marked in ellipses in Fig.~\ref{fig_CDFs_RoI_others}). Based on the new RoI
definitions, the mean, median, standard deviation of blockage loss and the RoI's coverage area
in the sphere are described in Table~\ref{table_blockage_RoIs}. From this table, we note that
${\cal R}_2$ and ${\cal R}_4$ describe more reflection gains with reduced values of blockage
loss (compared with ${\cal R}_1$). On the other hand, ${\cal R}_3$ and ${\cal R}_5$ are
comparable with ${\cal R}_1$. These behaviors are specific to the parameters used for
understanding blockage behavior (note that ${\cal R}_2$, ${\cal R}_3$ and ${\cal R}_4$ can
be made comparable by careful parameter tuning).

Note that ${\cal R}_1$ corresponds to the RoI with just Freespace information alone, whereas
${\cal R}_5$ corresponds to Freespace as well as hand/body information. In general, we are
interested in performing a head-to-head comparison between these two RoIs as they generally
capture how RoI is defined in prior works and how RoI can be modified by incorporating
blockage information. For this head-to-head comparison, we choose $\Delta_1$ and $\Delta_5$ to
ensure that the EIRP is below a fixed level in both cases (in this comparison, we choose $-35$,
$-40$ and $-45$ dBm as benchmarks). Table~\ref{table_blockage_RoIs} compares ${\cal R}_1$ with
${\cal R}_5$ and shows that ${\cal R}_5$ only improves the area of interest on an absolute scale
by $1\%$-$2.5\%$ (relatively from $2\%$-$4.5\%$) of the sphere. This is not a substantial increase in coverage area and the
reason for this small increase is that there are hardly any reflection gains in the hard hand
grip mode over any part of the sphere. Thus, in this case, {\em ${\cal R}_1$ is sufficient to
capture blockage performance}. However, we will see in Sec.~\ref{sec4} that ${\cal R}_5$ and other
RoIs become useful with looser hand grips.

\begin{table*}[htb!]
\caption{Blockage Performance with Different RoIs for the Hard Hand Grip (Study 1)}
\label{table_blockage_RoIs}
\begin{center}
\begin{tabular}{|c||c|c|c|c||}
\hline
Criterion & Mean (in dB) & Median (in dB) & Std.\ deviation (in dB) & Percentage of sphere \\
\hline
${\cal R}_1$, $\Delta_1 = 5$ dB & $14.9$ & $16.0$ & $7.3$ & $14.6$ \\ \hline
${\cal R}_1$, $\Delta_1 = 10$ dB & $14.7$ & $15.6$ & $8.4$ & $32.6$ \\ \hline
${\cal R}_2$, $\Delta_1 = \Delta_2 = 5$ dB & $12.6$ & $14.3$ & $9.3$ & $16.7$ \\ \hline
${\cal R}_2$, $\Delta_1 = 5$ dB, $\Delta_2 = 10$ dB & $8.9$ & $8.3$ & $10.3$ & $22.7$ \\ \hline
${\cal R}_3$, $\Delta_1 = \Delta_3 = 5$ dB & $14.4$ & $15.7$ & $8.1$ & $14.9$ \\ \hline
${\cal R}_3$, $\Delta_1 = 5$ dB, $\Delta_3 = 10$ dB & $11.0$ & $12.6$ & $9.9$ & $19.0$ \\ \hline
${\cal R}_4$, $\Delta_1 = 5$ dB, $\Delta_4 = -35$ dBm & $11.9$ & $13.7$ & $9.6$ & $17.7$ \\ \hline
${\cal R}_4$, $\Delta_1 = 5$ dB, $\Delta_4 = -40$ dBm & $8.8$ & $7.6$ & $10.3$ & $23.3$ \\ \hline
${\cal R}_4$, $\Delta_1 = 5$ dB, $\Delta_4 = -45$ dBm & $7.6$ & $6.1$ & $9.5$ & $31.2$ \\ \hline
${\cal R}_5$, $\Delta_5 = -35$ dBm & $13.9$ & $15.4$ & $9.2$ & $30.8$ \\ \hline
${\cal R}_5$, $\Delta_5 = -40$ dBm & $14.2$ & $15.4$ & $9.9$ & $57.2$ \\ \hline
${\cal R}_5$, $\Delta_5 = -45$ dBm & $13.7$ & $14.1$ & $9.6$ & $81.8$
\\ \hline \hline
\end{tabular}
\end{center}
%\end{table*}
%\begin{table*}
\begin{center}
\begin{tabular}{|c|| c|c|c|c|| c|c|c|c|| c|c||}
\hline
& \multicolumn{4}{c||}{${\cal R}_1$}
& \multicolumn{4}{c||}{${\cal R}_5$}
& \multicolumn{2}{c||}{Improvement}
\\ \cline{2-11}
EIRP & Mean  & Median & Std.\ dev.\ & $\%$ of sph. &
Mean & Median & Std.\ dev.\ & $\%$ of sph.  & Abs.\ & Rel.\
\\
(in dBm) & (in dB) & (in dB) & (in dB) & (in $\%$) & (in dB) & (in dB) & (in dB) &
(in $\%$) & (in $\%$) & (in $\%$) \\
\hline
$> -35$ & $14.7$ & $15.6$ & $8.3$ & $29.7$ & $13.9$ & $15.4$ & $9.2$ & $30.8$ & $1.1$ & $3.7$
\\ \hline
$> -40$ & $15.1$ & $15.7$ & $8.9$ & $54.7$ & $14.2$ & $15.4$ & $9.9$ & $57.2$ & $2.5$ & $4.6$
\\ \hline
$> -45$ & $14.1$ & $14.3$ & $9.1$ & $80.1$ & $13.7$ & $14.1$ & $9.6$ & $81.8$ & $1.7$ & $2.1$
\\ \hline \hline
\end{tabular}
%\end{center}
%{\vspace{0.1in}}
%%%%%%%%%%%%%%%%
%\begin{center}
\end{center}
\end{table*}

\section{Beamforming Performance of Other Subarrays and Hand Holdings}
\label{sec4}
We now present results from Studies 2-5 which cover other subarrays and hand holdings.

\subsection{Study 2: $4 \times 1$ Patch Subarray with a Loose Hand Grip}
\label{sec4a}
We start with the same $4 \times 1$ patch subarray in Module 3 (as in Study 1), but with a left hand
holding which ensures that only a few fingers cover the antennas in the module. As a result,
we see blockage performance with an essentially loose hand grip mode.

\begin{figure*}[htb!]
\begin{center}
\begin{tabular}{cc}
\includegraphics[height=1.9in,width=2.8in]{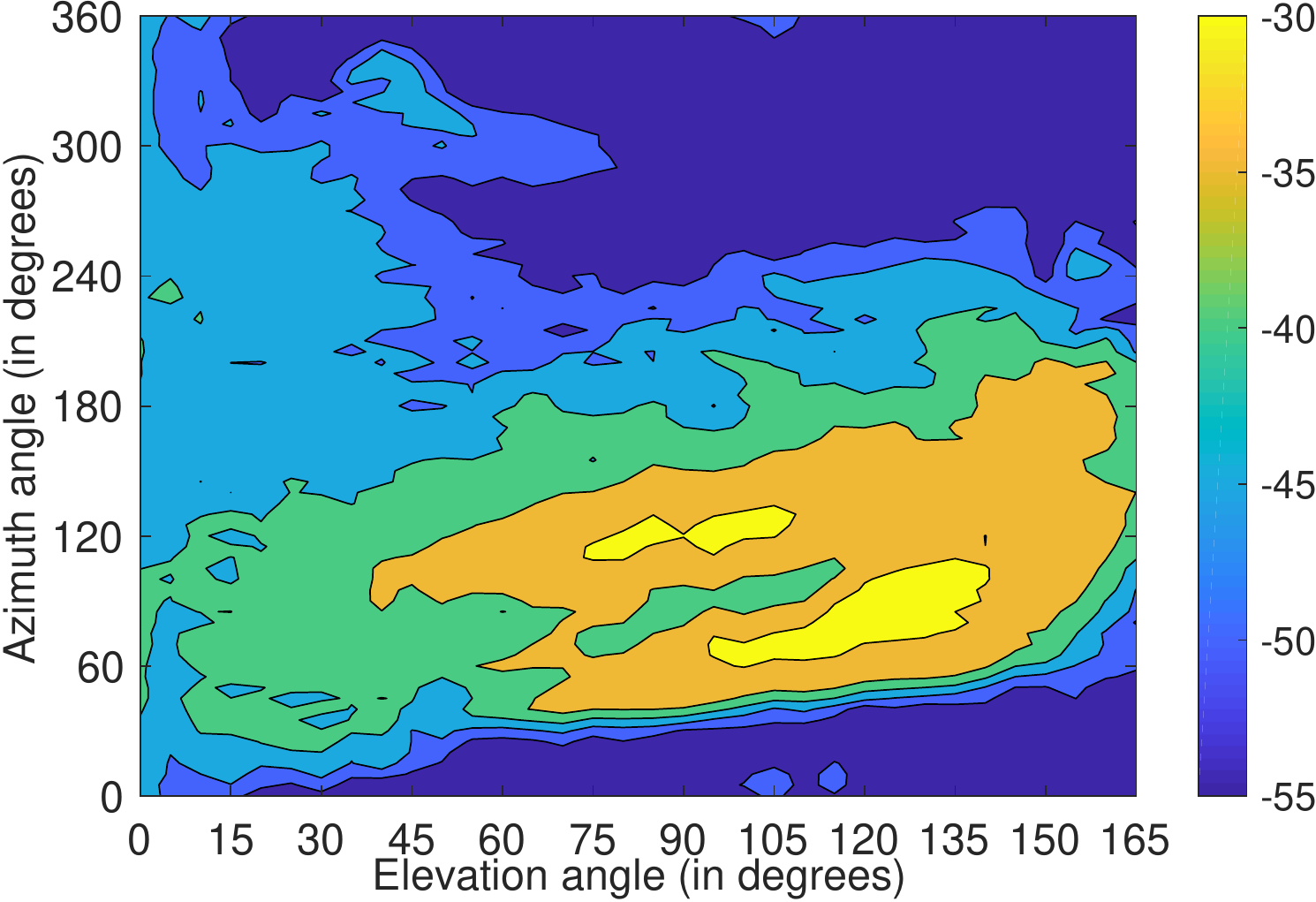}
&
\includegraphics[height=1.9in,width=2.8in]{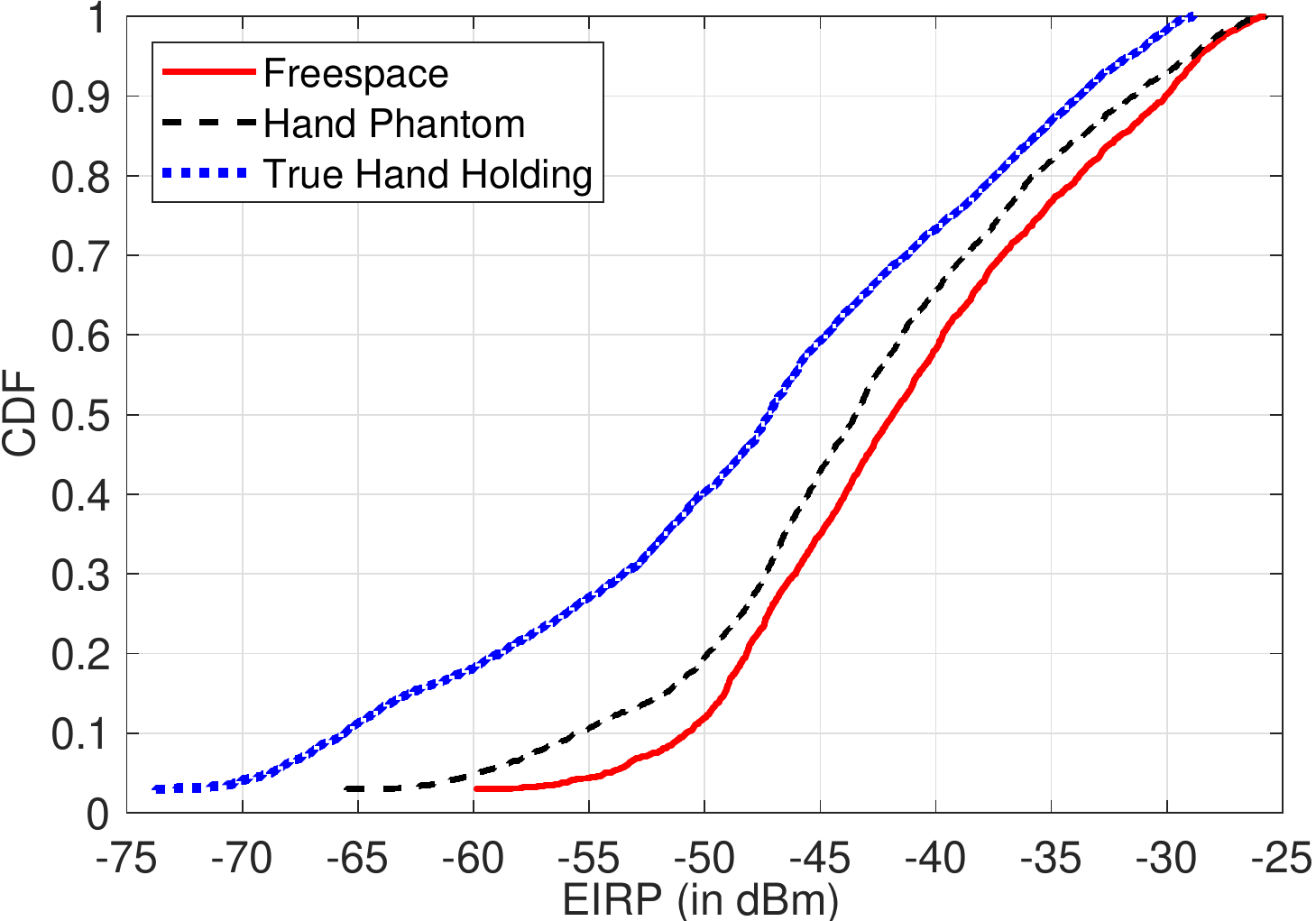}
\\
(a) & (b)
\end{tabular}
\caption{\label{fig_Study2}
(a) Overlay plot of the best of the three beam patterns in Freespace with the $4 \times 1$ patch subarray
in loose hand grip mode. (b) CDFs of EIRP in the $4 \times 1$ patch subarray with loose hand grip. }
\end{center}
%\vspace{-5mm}
\end{figure*}

Fig.~\ref{fig_Study2}(a) illustrates the overlay plot of the best of the three beams
used with this subarray under the loose hand grip mode. The close similarity between
this plot and the Freespace plot in Fig.~\ref{fig_overlay_4by1patch}(a) suggests that the
impact of the loose hand grip on blockage loss is significantly smaller than in the hard
hand grip mode. Reflecting this observation, the CDF comparison of EIRPs in Fig.~\ref{fig_Study2}(b)
and Table~\ref{table_blockage_Study2345} show that $10\%$-$25\%$ of the spherical coverage
is lost at different EIRP levels, or an equivalent $3.5$-$10.5$ dB loss at different
spherical coverage levels. On the other hand, the hand phantom shows only a $5\%$-$8\%$ loss of
spherical coverage, which is grossly mismatched to the observations with the true hand. The
corresponding numbers for the hard hand grip are $20\%$-$45\%$ and $8.5$-$17$ dB, respectively,
showing that the hand grip has a {\em significant} impact on the blockage loss observed. To
understand the efficacy of an enhanced blockage region such as ${\cal R}_5$, we perform the
same head-to-head comparison between these two RoIs as in the hard hand grip case.
Table~\ref{table_RoIs_Study2345} shows that (unlike the hard hand grip case where {\em only}
$1\%$-$2.5\%$ improvement in spherical coverage was observed with ${\cal R}_5$ over ${\cal R}_1$)
we observe a $2\%$-$6.5\%$ improvement in absolute spherical coverage with ${\cal R}_5$, which
is substantial. The primary reasons for this enhancement are the gains with hand reflections which
are captured by ${\cal R}_5$, but not with ${\cal R}_1$, illustrating the need for such an enhanced
RoI definition.

\begin{table*}[htb!]
\caption{EIRP-Based Comparisons for Other Studies}
\label{table_blockage_Study2345}
\begin{center}
\begin{tabular}{|l|c| c| c|c|c| c|c|c|| c|c|c||}
\hline
\multirow{5}[6]{0.2cm}{\rotatebox[origin=c]{90}{Study 2}} &
& \multicolumn{7}{c||}{Spherical Coverage Lost}
& \multicolumn{3}{c||}{Loss (in dB)} \\ \cline{2-12}
& EIRP & Freespace
& \multicolumn{3}{c|}{Hand Phantom (in $\%$)}
& \multicolumn{3}{c||}{True Hand (in $\%$)}
& Percentile & Hand Phantom & True Hand
\\ \cline{4-12}
& (in dBm) & (in $\%$) & & Abs. & Rel. & & Abs. & Rel. & $90$ & $1.4$ & $3.7$
\\ \cline{2-12}
& $>-35$ & $23.3$ & $18.1$ & $5.1$ & $22.0$ & $13.1$ & $10.2$ & $43.9$ & $80$ & $2.0$ & $3.6$ \\
\cline{2-12}
%%%
& $>-40$ & $41.8$ & $34.4$ & $7.4$ & $17.7$ & $26.8$ & $15.0$ & $36.0$ & $50$ & $1.6$ & $5.4$ \\
\cline{2-12}
%%%
& $>-45$ & $65.0$ & $57.2$ & $7.8$ & $12.0$ & $40.7$ & $24.3$ & $37.4$ & $20$ & $1.7$ & $10.6$ \\
\hline %\hline
%%%%%%%%%%%%%%%%%%%%%%
%%%%%%%%%%%%%%%%%%%%%%
%%%%%%%%%%%%%%%%%%%%%%
\hline
\multirow{4}[0]{0.2cm}{\rotatebox[origin=c]{90}{Study 3}}
%& EIRP & Freespace
%& \multicolumn{3}{c|}{Hand Phantom (in $\%$)}
%& \multicolumn{3}{c||}{True Hand (in $\%$)}
& \multicolumn{8}{c||}{}
& $90$ & $2.6$ & $15.9$ \\
\cline{2-12}
& $>-40$ & $35.3$ & $22.1$ & $13.2$ & $37.3$ & $0.0$ & $35.3$ & $100.0$ & $80$ & $2.7$ & $16.3$ \\
\cline{2-12}
%%%
& $>-45$ & $58.9$ & $46.4$ & $12.5$ & $21.3$ & $0.7$ & $58.2$ & $98.8$ & $50$ & $2.6$ & $15.9$ \\
\cline{2-12}
%%%
& $>-50$ & $84.1$ & $69.9$ & $14.1$ & $16.8$ & $7.7$ & $76.4$ & $90.8$ & $20$ & $4.2$ & $19.7$ \\
\hline
%%%%%%%%%%%%%%%%%%%%%%
%%%%%%%%%%%%%%%%%%%%%%
%%%%%%%%%%%%%%%%%%%%%%
\hline
\multirow{4}[0]{0.2cm}{\rotatebox[origin=c]{90}{Study 4}}
%& EIRP & Freespace
%& \multicolumn{3}{c|}{Hand Phantom (in $\%$)}
%& \multicolumn{3}{c||}{True Hand (in $\%$)}
& \multicolumn{8}{c||}{}
& $90$ & $0.4$ & $0.4$ \\
\cline{2-12}
&
$>-40$ & $35.3$ & $35.3$ & $0.0$ & $0.0$ & $26.2$ & $9.1$ & $25.7$ & $80$ & $0.2$ & $1.5$ \\
\cline{2-12}
%%%
& $>-45$ & $58.9$ & $61.5$ & $-2.5$ & $-4.4$ & $47.1$ & $11.8$ & $20.0$ & $50$ & $-0.4$ & $2.6$ \\
\cline{2-12}
%%%
& $>-50$ & $84.1$ & $79.6$ & $4.5$ & $5.3$ & $64.5$ & $19.5$ & $23.2$ & $20$ & $1.5$ & $10.8$ \\
\hline
%%%%%%%%%%%%%%%%%%%%%%
%%%%%%%%%%%%%%%%%%%%%%
%%%%%%%%%%%%%%%%%%%%%%
\hline
\multirow{4}[0]{0.2cm}{\rotatebox[origin=c]{90}{Study 5}}
& \multicolumn{8}{c||}{}
& $90$ & $10.5$ & $12.7$ \\
\cline{2-12}
&
$>-45$ & $28.7$ & $2.7$ & $26.1$ & $90.7$ & $0.04$ & $28.7$ & $99.9$ & $80$ & $10.8$ & $12.4$ \\
\cline{2-12}
%%%
& $>-50$ & $50.3$ & $11.7$ & $38.6$ & $76.8$ & $7.6$ & $42.8$ & $85.0$ & $50$ & $10.7$ & $10.9$ \\
\cline{2-12}
%%%
& $>-55$ & $70.7$ & $26.5$ & $44.2$ & $62.5$ & $22.2$ & $48.5$ & $68.6$ & $20$ & $9.8$ & $9.5$ \\
\hline
\end{tabular}
%\end{center}
%{\vspace{0.1in}}
%%%%%%%%%%%%%%%%
%\begin{center}
\end{center}
\end{table*}
%\end{center}

\begin{table*}
\caption{Blockage Performance of ${\cal R}_1$ and ${\cal R}_5$ for Other Studies}
\label{table_RoIs_Study2345}
\begin{center}
\begin{tabular}{|l|c|| c|c|c|c|| c|c|c|c|| c|c||}
\hline
\multirow{6}[6]{0.2cm}{\rotatebox[origin=c]{90}{Study 2}} &
& \multicolumn{4}{c||}{${\cal R}_1$}
& \multicolumn{4}{c||}{${\cal R}_5$}
& \multicolumn{2}{c||}{Improvement}
\\ \cline{3-12}
& EIRP & Mean  & Median & Std.\ dev.\ & $\%$ of sph. &
Mean & Median & Std.\ dev.\ & $\%$ of sph.  & Abs.\ & Rel.\
\\
& (in dBm) & (in dB) & (in dB) & (in dB) & (in $\%$) & (in dB) & (in dB) & (in dB) &
(in $\%$) & (in $\%$) & (in $\%$) \\
\cline{2-12}
& $> -35$ & $5.1$ & $4.1$ & $4.6$ & $29.7$ & $4.2$ & $3.8$ & $5.9$ & $31.7$ & $2.0$ & $6.7$
\\ \cline{2-12}
& $> -40$ & $7.5$ & $5.3$ & $7.5$ & $54.7$ & $6.4$ & $4.7$ & $8.6$ & $58.4$ & $3.7$ & $6.8$
\\ \cline{2-12}
& $> -45$ & $7.3$ & $4.9$ & $8.1$ & $80.1$ & $5.8$ & $4.3$ & $9.4$ & $86.5$ & $6.4$ & $8.0$
\\ \hline %\hline
%%%%%%%%%%%%%%%%%%%%%%
%%%%%%%%%%%%%%%%%%%%%%
%%%%%%%%%%%%%%%%%%%%%%
\hline
%\multirow{6}[6]{0.2cm}{\rotatebox[origin=c]{90}{Study 3}} &
%& \multicolumn{4}{c||}{${\cal R}_1$}
%& \multicolumn{4}{c||}{${\cal R}_5$}
%& \multicolumn{2}{c||}{Improvement}
%\\ \cline{3-12}
%\multirow{5}[6]{0.2cm}{\rotatebox[origin=c]{90}{Study 3}}
%& EIRP & Mean  & Median & Std.\ dev.\ & $\%$ of sph. &
%Mean & Median & Std.\ dev.\ & $\%$ of sph.  & Abs.\ & Rel.\
%\\
%& (in dBm) & (in dB) & (in dB) & (in dB) & (in $\%$) & (in dB) & (in dB) & (in dB) &
%(in $\%$) & (in $\%$) & (in $\%$) \\
%\cline{2-12}
\multirow{3}[0]{-0.2cm}{\rotatebox[origin=c]{90}{Study 3}}
& $> -40$ & $19.3$ & $18.4$ & $8.1$ & $51.3$ & $19.3$ & $18.4$ & $8.1$ & $51.3$ & $0$ & $0$
\\ \cline{2-12}
& $> -45$ & $18.6$ & $18.2$ & $8.0$ & $72.7$ & $18.6$ & $18.2$ & $8.0$ & $72.7$ & $0$ & $0$
\\ \cline{2-12}
& $> -50$ & $17.7$ & $17.6$ & $8.0$ & $87.1$ & $17.2$ & $17.3$ & $9.0$ & $88.6$ & $1.5$ & $1.7$
\\ \hline %\hline
%%%%%%%%%%%%%%%%%%%%%%
%%%%%%%%%%%%%%%%%%%%%%
%%%%%%%%%%%%%%%%%%%%%%
\hline
\multirow{3}[0]{-0.2cm}{\rotatebox[origin=c]{90}{Study 4}}
& $> -40$ & $3.7$ & $0.4$ & $7.5$ & $51.3$ & $1.5$ & $0.04$ & $10.4$ & $57.9$ & $6.6$ & $12.9$
\\ \cline{2-12}
& $> -45$ & $4.8$ & $0.8$ & $8.3$ & $72.7$ & $2.9$ & $0.4$ & $10.5$ & $80.5$ & $7.8$ & $10.7$
\\ \cline{2-12}
& $> -50$ & $5.1$ & $1.3$ & $8.4$ & $87.2$ & $3.4$ & $0.7$ & $10.4$ & $94.6$ & $7.4$ & $8.5$
\\ \hline
%%%%%%%%%%%%%%%%%%%%%%
%%%%%%%%%%%%%%%%%%%%%%
%%%%%%%%%%%%%%%%%%%%%%
\hline
\multirow{3}[0]{-0.2cm}{\rotatebox[origin=c]{90}{Study 5}}
& $> -45$ & $18.3$ & $18.4$ & $5.6$ & $39.5$ & $15.9$ & $17.9$ & $9.5$ & $43.0$ & $3.5$ & $8.9$
\\ \cline{2-12}
& $> -50$ & $14.9$ & $16.6$ & $7.8$ & $63.9$ & $11.3$ & $15.5$ & $11.8$ & $73.9$ & $10.0$ & $15.7$
\\ \cline{2-12}
& $> -55$ & $13.1$ & $14.6$ & $8.4$ & $81.8$ & $10.5$ & $13.6$ & $11.3$ & $90.8$ & $9.0$ & $11.0$
\\ \hline
\end{tabular}
%\end{center}
%{\vspace{0.1in}}
%%%%%%%%%%%%%%%%
%\begin{center}
\end{center}
\end{table*}

\begin{figure*}[htb!]
\begin{center}
\begin{tabular}{cc}
\includegraphics[height=1.9in,width=2.8in]{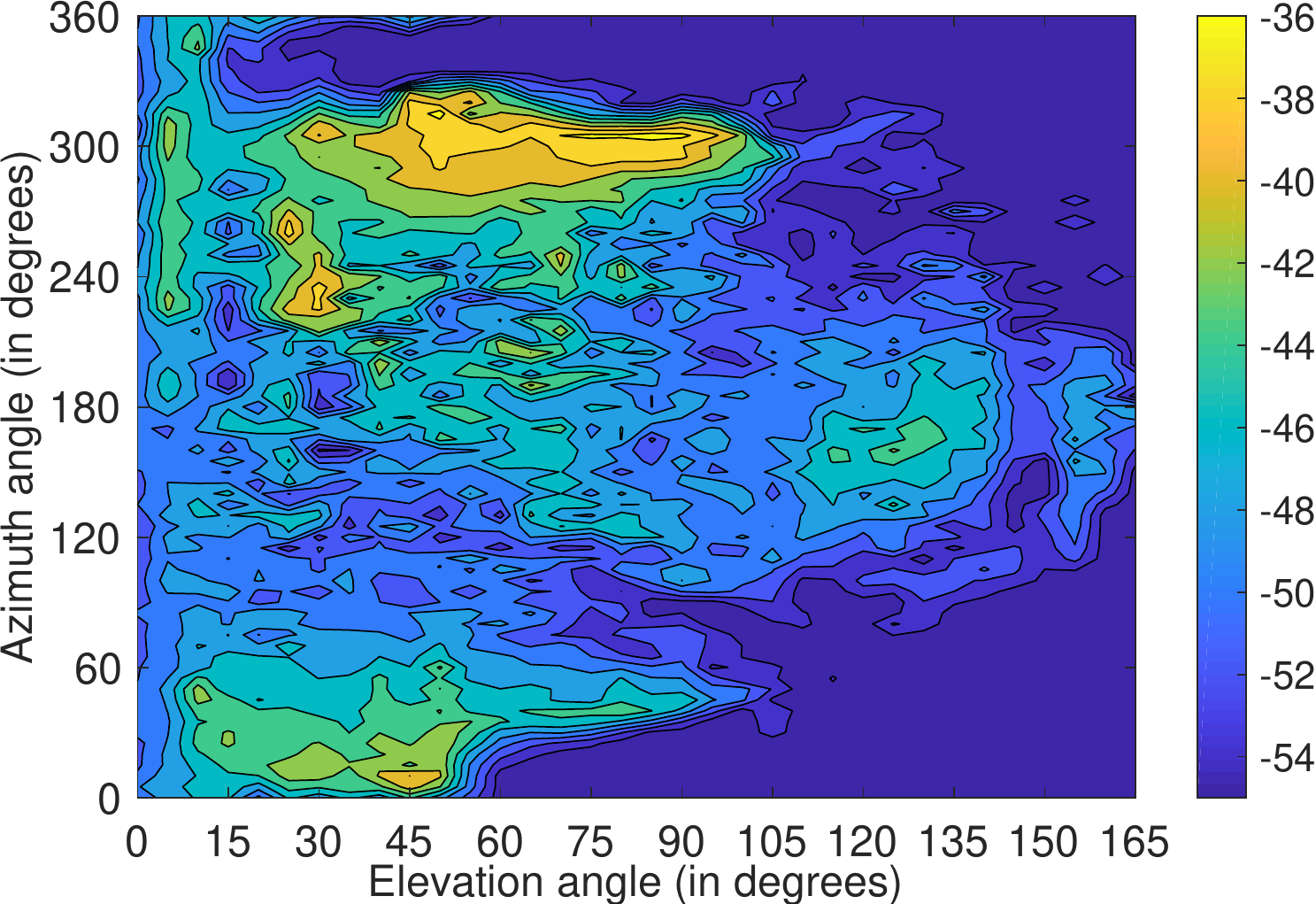}
&
\includegraphics[height=1.9in,width=2.8in]{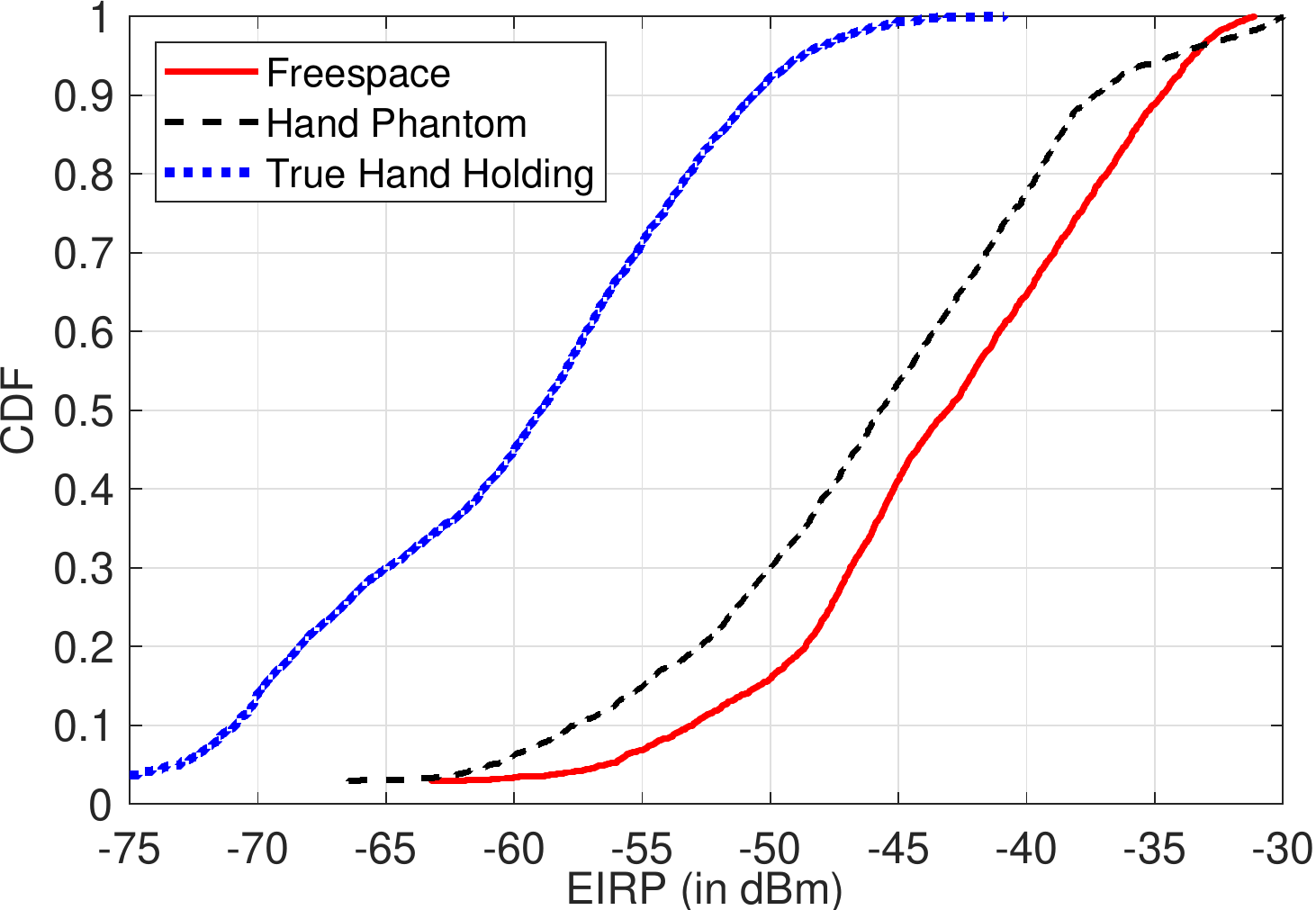}
\\
(a) & (d) \\
\includegraphics[height=1.9in,width=2.8in]{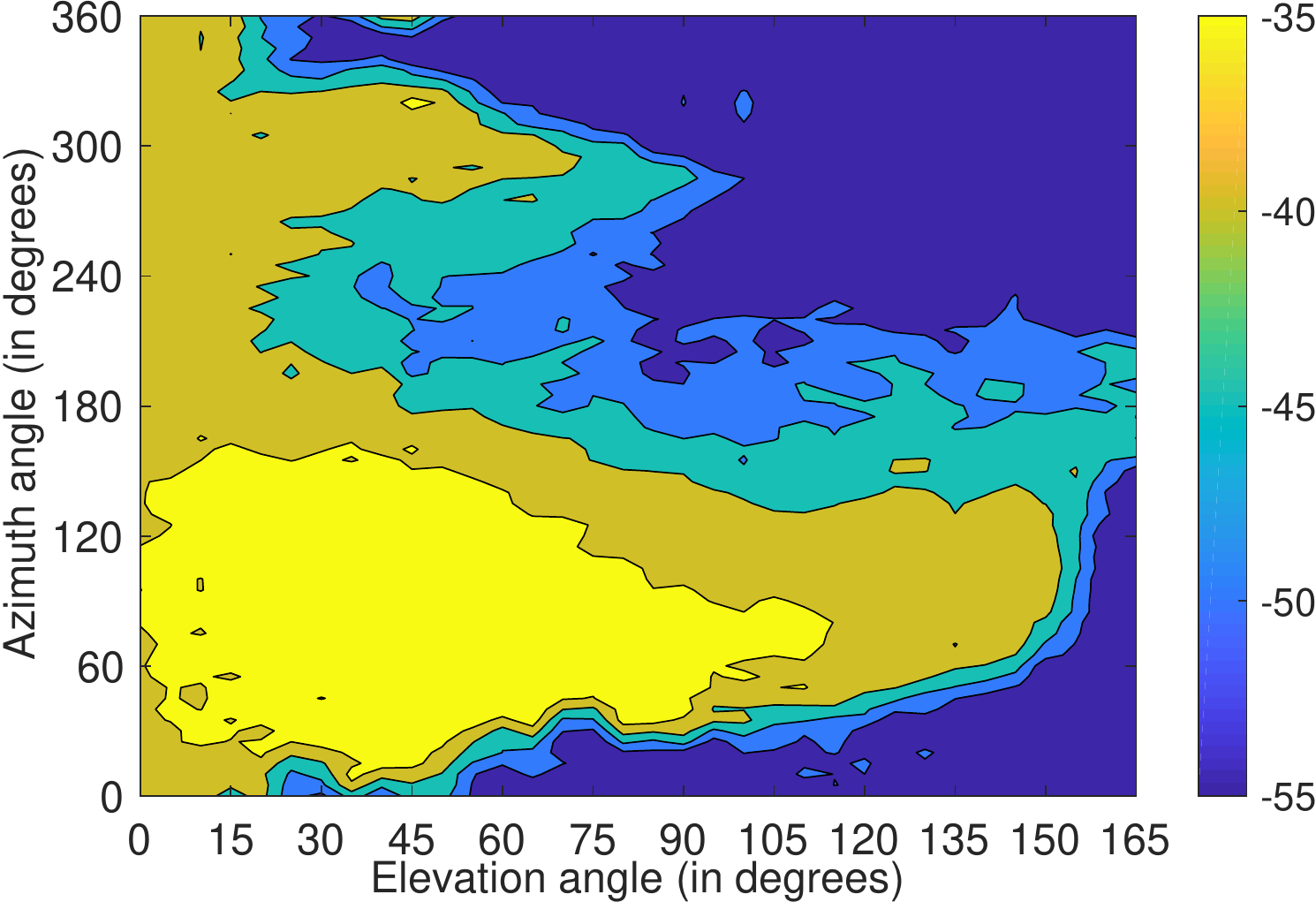}
&
\includegraphics[height=1.9in,width=2.8in]{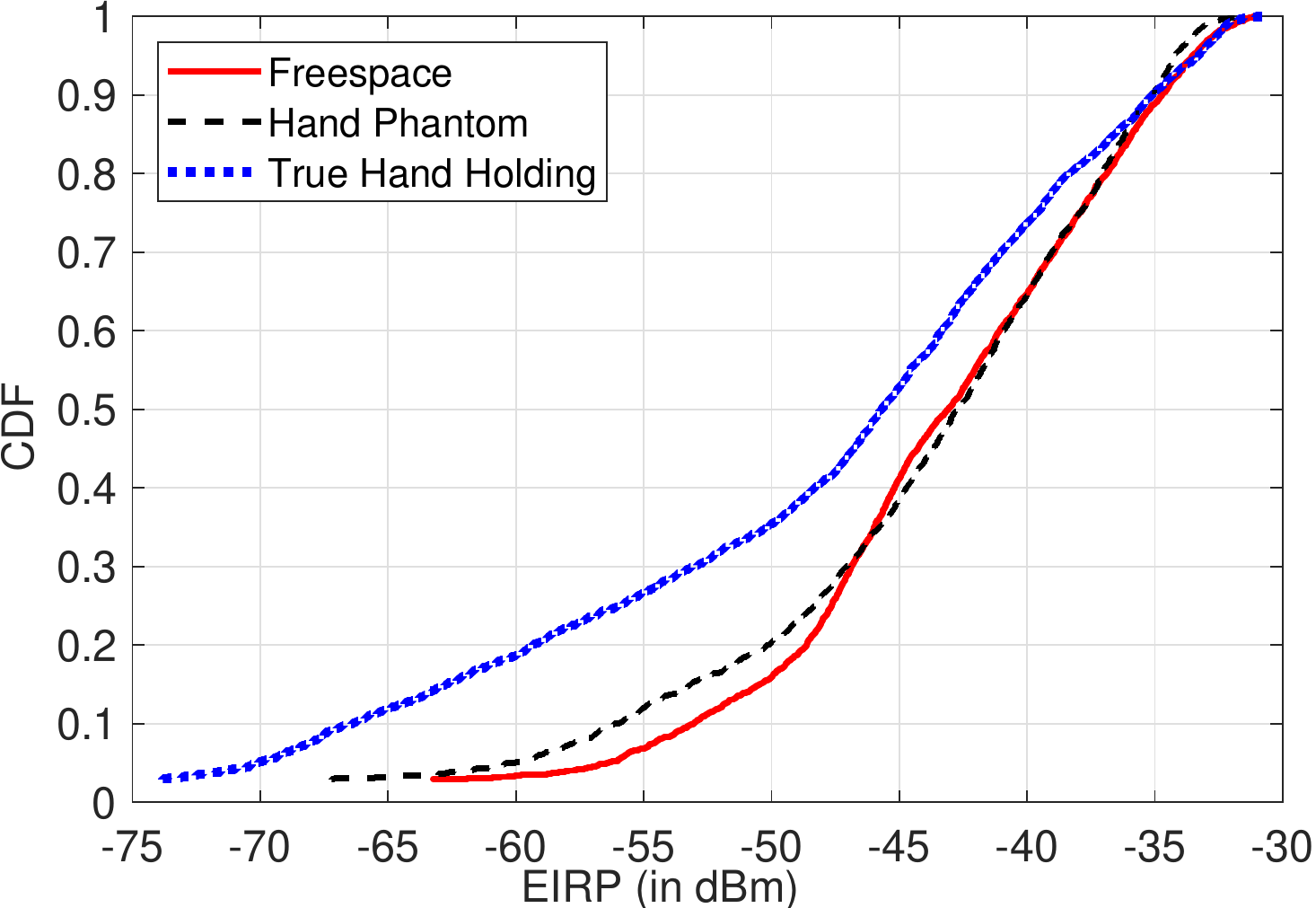}
\\
(b) & (e)
\\
\includegraphics[height=1.9in,width=2.8in]{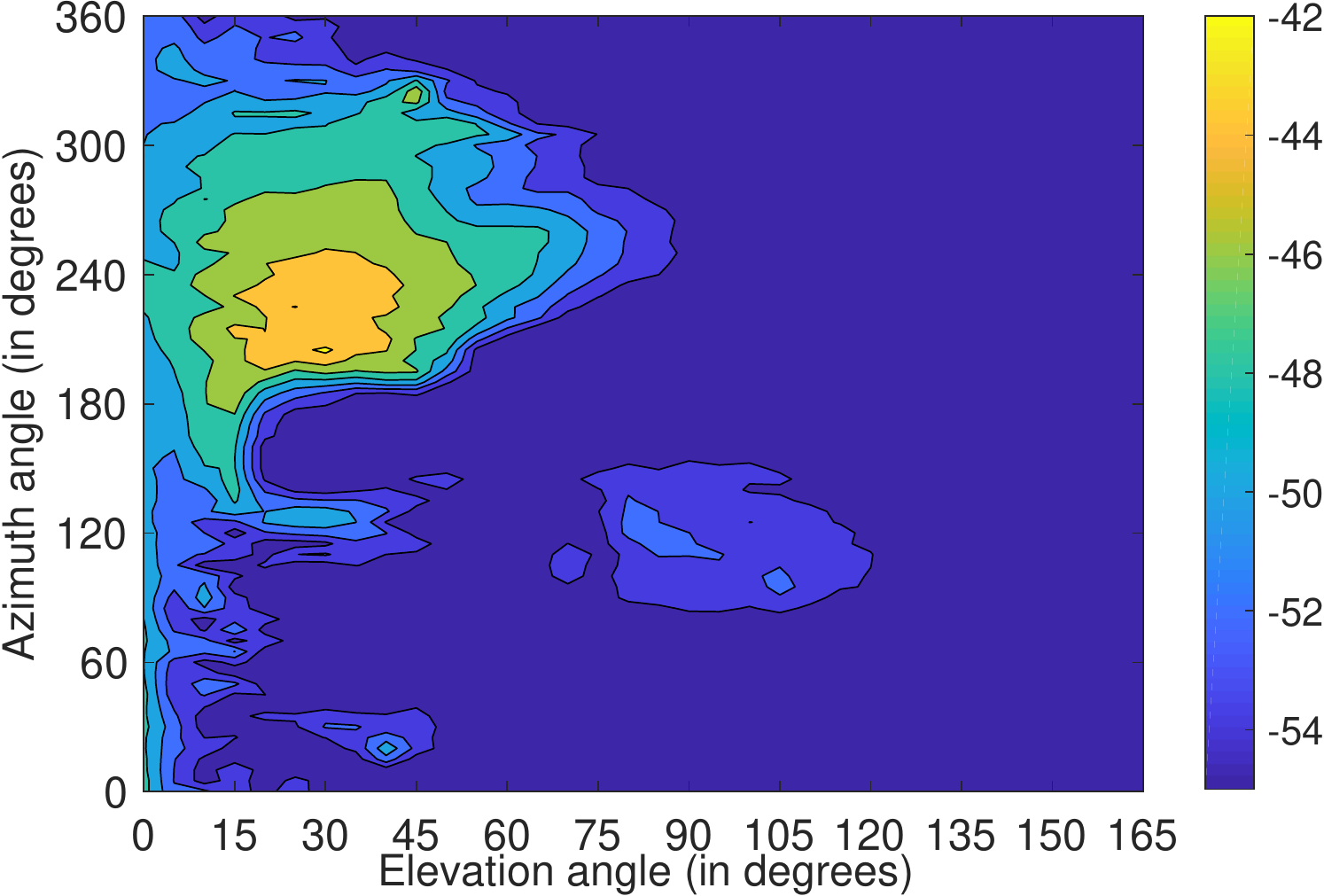}
&
\includegraphics[height=1.9in,width=2.8in]{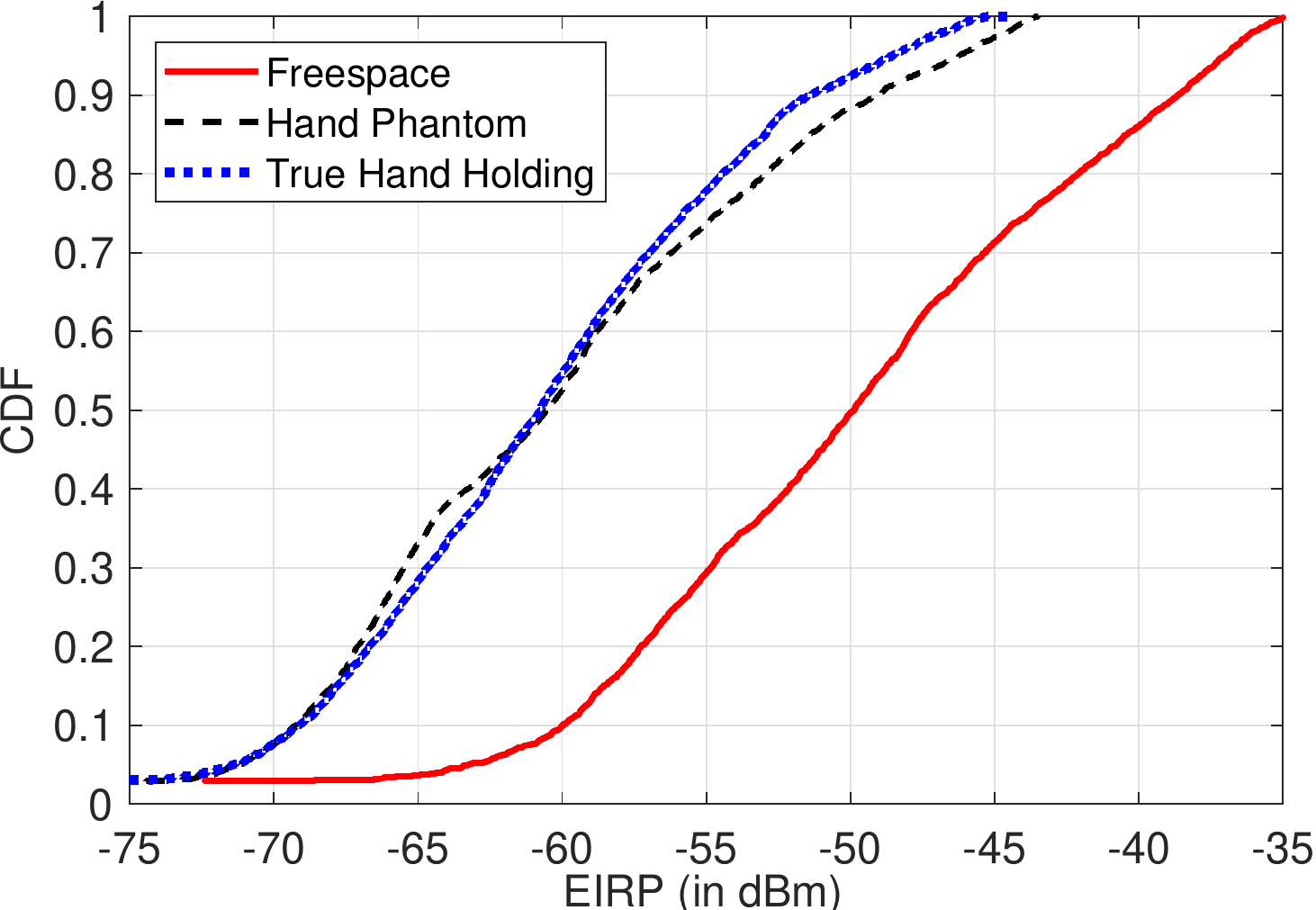}
\\
(c) & (f)
\end{tabular}
\caption{\label{fig_Study345}
(a)-(c) Overlay plot of the best of the three beam patterns in Freespace for Studies 3-5.
(d)-(f) CDFs of EIRP with different hand modes in Studies 3-5. }
\end{center}
%\vspace{-5mm}
\end{figure*}

\subsection{Study 3: $2 \times 1$ Dipole Subarray with a Hard Hand Grip}
\label{sec4b}
We now consider the beamformed performance of the $2 \times 1$ dipole array (in Module 3)
with a hard
hand grip. For the dipoles, we again consider a codebook of three beams steering energy
towards the boresight of the array, $+45^{\sf o}$ to the boresight and $-45^{\sf o}$ to
the boresight, respectively. The change from $\pm 30^{\sf o}$ for the $4 \times 1$ subarray
to a $\pm 45^{\sf o}$ for the $2 \times 1$ subarray is due to beamwidth differences for
different-sized arrays. %The beam patterns of these three beams in Freespace are illustrated
%in Figs.~\ref{fig_beampatterns_2by1dipole}(a)-(c), respectively. Note the regularity of the
%beam patterns and the beamwidth of each beam is $\approx 40^{\sf o}$-$45^{\sf o}$ (which
%is as expected for a $2 \times 1$ subarray~\cite{balanis}). The beam patterns with the
%hand phantom and a true hand are then plotted in Figs.~\ref{fig_beampatterns_2by1dipole}(d)-(i),
%respectively. These beam patterns show the significant distortion introduced by the hand
%and the poor reproducibility of the hand behavior by the hand phantom.
The beam patterns of these three beams in Freespace appear regular (not illustrated here
pictorially due to space constraints) with the beamwidth of each beam being
$\approx 40^{\sf o}$-$45^{\sf o}$ (which is as expected for a $2 \times 1$ subarray~\cite{balanis}).
The beam patterns with the hand phantom and a true hand show significant distortions
(again not illustrated pictorially) introduced by the hand and the poor reproducibility of
the hand behavior by the hand phantom.

Reflecting these observations, the overlay plot of the beam patterns of the three best beams
is plotted in Fig.~\ref{fig_Study345}(a), which shows a small ocean of yellow (or high
signal strength regions). In comparison, the CDFs of the EIRPs in Fig.~\ref{fig_Study345}(b)
shows an almost constant gap of $\approx 15$ dB between the Freespace and true hand holding
performance. More precisely, Table~\ref{table_blockage_Study2345} shows that $35\%$-$75\%$ of
the spherical coverage is lost at different EIRP levels, or an equivalent $15$-$20$ dB loss
at different spherical coverage levels. As with the patches case, ${\cal R}_5$ does not seem
to bring in any additional value over ${\cal R}_1$ showing that hand reflections are not
important with the hard hand grip.

\subsection{Study 4: $2 \times 1$ Dipole Subarray with a Loose Hand Grip}
\label{sec4c}
In the loose hand grip mode, the $2 \times 1$ dipole subarray shows a similar behavior
as the $4 \times 1$ patch subarray with a loose hand grip (Study 2). At different EIRP levels,
a $9\%$-$20\%$ spherical coverage is lost relative to the Freespace case corresponding
to a $0$-$11$ dB blockage loss at different spherical coverage levels. The CDF with the
hand phantom is similar to the Freespace scenario showing the poor match of the hand
phantom in capturing true blockage performance. In contrast to the hard hand grip case,
we see that ${\cal R}_5$ can result in substantial coverage improvement over ${\cal R}_1$
(of $6.5\%$-$8\%$ absolute improvement or $8.5\%$-$13\%$ relative improvement). This
study shows that hand reflections with loose hand grips need to be carefully captured
with RoI such as ${\cal R}_5$ and a RoI such as ${\cal R}_1$ is {\em not} sufficient
in such scenarios.

\subsection{Study 5: $4 \times 1$ Patch Subarray with an Intermediate Hand Grip}
\label{sec4d}
In the final study, a two-handed grip in the Landscape mode over the $4 \times 1$ patch subarray
is considered. This study shows some effects observed with the hard hand grip as well
as some effects observed with the loose hand grip studies. In terms of EIRP losses, we see
a loss of $9.5$-$13$ dB at different spherical coverage levels, which is equivalent to a
spherical coverage loss of $25\%$-$50\%$ at different EIRP levels. This observation is
similar to those made in the hard hand grip cases (Studies 1 and 3). On the other hand,
mirroring Studies 2 and 4 (loose hand grip cases),
${\cal R}_5$ leads to a substantial improvement over ${\cal R}_1$ of $3.5\%$-$10\%$ absolute
spherical coverage increase with hand reflections (corresponding to a relative improvement
of $9\%$-$16\%$). Thus, in this study, we see that the hand position/grip leads to a
substantial performance decrease over the intended coverage of the subarray, but also a
substantial performance increase over other regions of the sphere where the subarray is not
intended to cover. Such aspects of blockage need to be carefully considered in understanding
the implications of blockage in practical settings.

\begin{table*}[htb!]
\caption{Statistics of Blockage Loss in the Five Studies}
\label{table_CDF_blockage_loss}
\begin{center}
\begin{tabular}{|c||c|c|c|c|c||}
\hline
Study & $\Delta_5$ (in dBm) & Mean (in dB) & Median (in dB) &
Std.\ deviation (in dB) & Percentage of sphere \\
\hline
1 & $-35$ & $13.9$ & $15.4$ & $9.2$ & $30.8\%$ \\ \hline
2 & $-35$ & $4.2$ & $3.8$ & $5.9$ & $31.7\%$ \\ \hline
3 & $-40$ & $19.3$ & $18.4$ & $8.1$ & $51.3\%$ \\ \hline
4 & $-40$ & $1.5$ & $0.04$ & $10.4$ & $57.9\%$ \\ \hline
5 & $-45$ & $15.9$ & $17.9$ & $9.5$ & $43.0\%$ \\
\hline \hline
\end{tabular}
\end{center}
\end{table*}

\section{Models for Blockage and its Impact on Physical Layer Performance}
\label{sec5}
We now explore good stochastic model fits for signal strength changes with hand/body
blockage and what these models imply for physical layer performance.

\begin{figure*}[htb!]
\begin{center}
\begin{tabular}{cc}
\includegraphics[height=1.9in,width=2.8in]{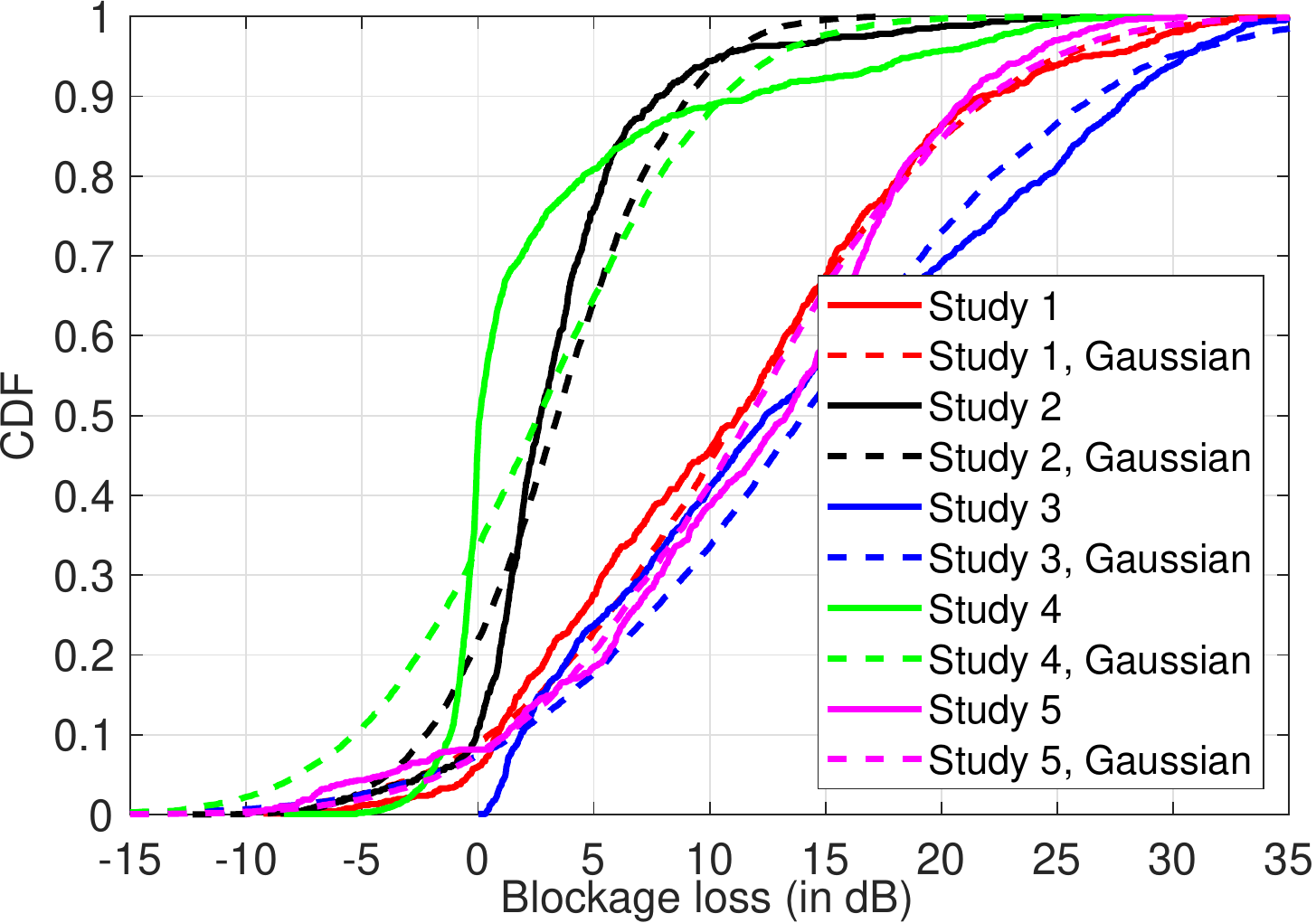}
&
\includegraphics[height=1.9in,width=2.8in]{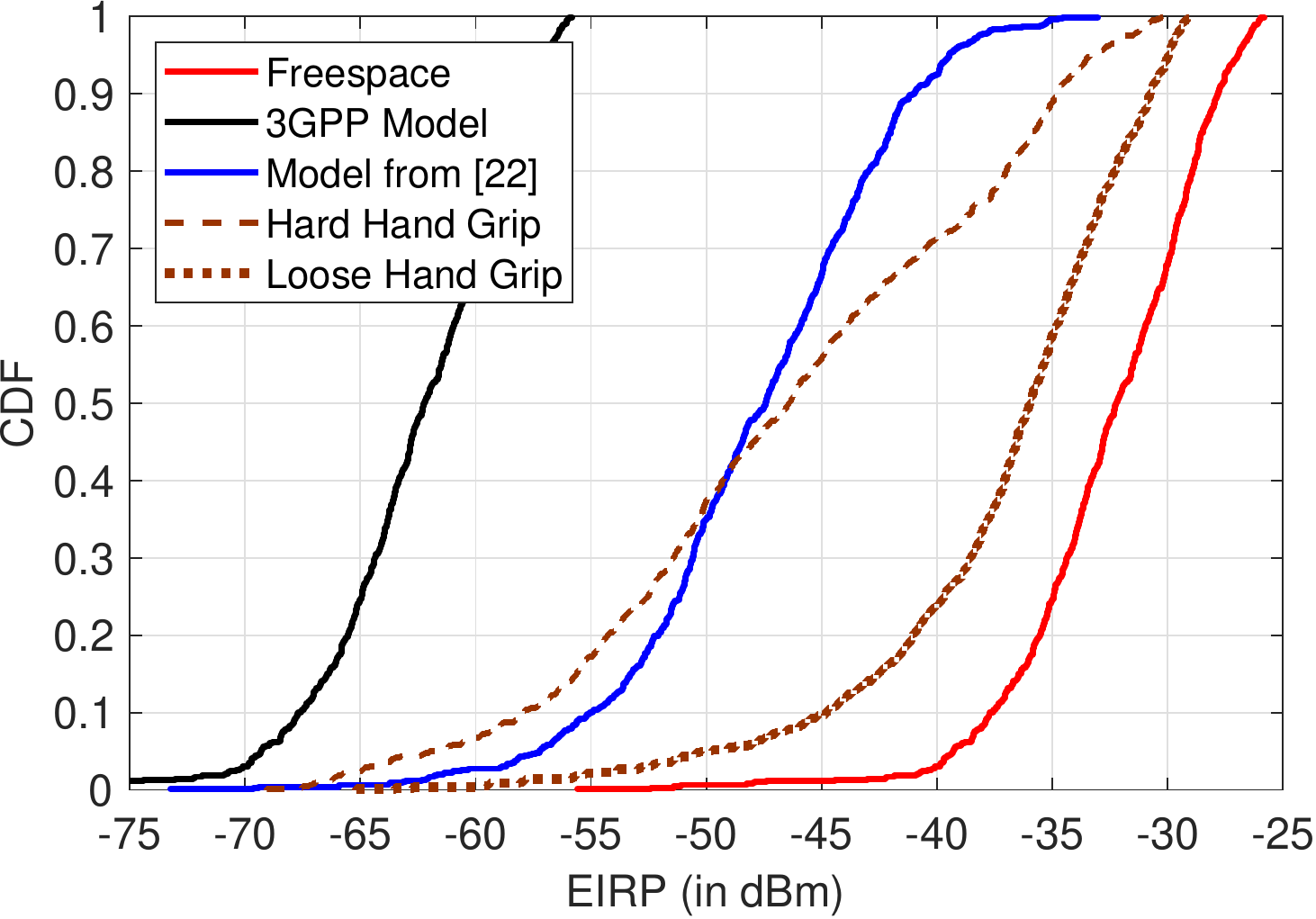}
\\
(a) & (b)
\end{tabular}
\caption{\label{fig_blockage_loss}
(a) CDF of blockage loss with ${\cal R}_5$ in all the five studies.
(b) EIRP comparison over ${\cal R}_5$ RoI with Freespace and different blockage models.}
\end{center}
%\vspace{-5mm}
\end{figure*}

Towards this goal, Fig.~\ref{fig_blockage_loss}(a) first plots the CDF of blockage loss
defined as the signal strength difference between Freespace and true hand holding scenarios
in the five studies with ${\cal R}_5$ leading to the RoI in these studies. To augment
Fig.~\ref{fig_blockage_loss}(a), Table~\ref{table_CDF_blockage_loss} illustrates the
statistics of blockage loss in these five studies along with the parameters that go into
the RoI definitions. In addition to plotting the empirical loss data, a simple Gaussian fit
with the mean and standard deviation of the data is also plotted for each of the five studies
in Fig.~\ref{fig_blockage_loss}(a). As mentioned earlier,
we note that the mean of blockage loss is substantially less in all the five cases (even
with the hard hand grip) relative to prior works that reported loss often in excess of
$30$ dB. The sources of discrepancies for such wide variations could include beamwidth
differences between phased array of antennas in commercial form-factor UEs
relative to horn antenna studies that have been reported in prior works, material property
differences between UEs and horns, reflections due to hand that is often unaccounted for
in prior works, etc. Fig.~\ref{fig_blockage_loss}(a) also shows that while simple Gaussian
models are good for hard/intermediate hand grips with substantial losses, looser hand grips
with a steeper loss curve and wider tails need more sophisticated multi-parameter
models such as Gamma
distribution, Weibull distribution, etc~\cite{vasanth_blockage_tap2018}. Empirical fits of
such distributions to data is the subject of ongoing work and will be reported elsewhere.

We now study the implications of blockage loss from true hand holding on beamforming performance
relative to models from prior works. For this, Fig.~\ref{fig_blockage_loss}(b) plots the EIRP
distribution seen over the Freespace scenario and the true hand holding in Studies 1 and 2 (hard
hand grip vs.\ loose hand grip mode) with the ${\cal R}_5$ RoI and by setting $\Delta_5 = -35$
dBm. Also, plotted are the EIRP deteriorations due to the 3GPP model~\cite{3gpp_CM_rel14_38901}
and the model from~\cite{vasanth_blockage_tap2018} for this RoI. Clearly, we note that the 3GPP
model widely over-estimates the blockage loss even in the hard hand grip case. On the other hand, the
model from~\cite{vasanth_blockage_tap2018} has a comparable performance to the hard hand grip
case, whereas it over-estimates the blockage loss in the loose hand grip case. Even within the
hard hand grip case, the model from~\cite{vasanth_blockage_tap2018} does not capture the hand
reflections and thus there is a cross-over between the CDFs observed here and the model
from~\cite{vasanth_blockage_tap2018} (better true performance at peak coverage points and weaker
true performance at lower tails). Such a cross-over can lead to a poor estimation of EIRP
(and thus physical layer performance) in a practical context, which requires careful study such
as the one in this work.

\section{Concluding Remarks}
\label{sec6}
The focus of this paper has been on understanding hand/body blockage with commercial
quality phased arrays in a user equipment operating at $28$ GHz. For this, a number of
controlled studies were performed and the impact of blockage was estimated with hard,
intermediate and loose hand grips. Our studies showed that blockage produces a complex
effect on the received signal strength depending on the direction of interest. In the main
scenario (also addressed in prior works), blockage leads to signal strength deterioration.
But unlike estimates from prior works which are excessive, we show that this deterioration
is moderate ($< 5$ dB for loose hand grip) to reasonable ($< 15$ dB for intermediate to hard
hand grips). Additionally, with looser hand grips and based on the hand holding, signals
can be reflected by the fingers, palm and different parts of the hand to improve signal strengths
in hitherto weak signal directions (as seen from a Freespace perspective). Such a complicated
behavior has not been explored or illustrated in prior works and this work documents such hand
reflection gains.

In terms of future work, this paper exposes the need for further careful studies in understanding
how blockage can affect millimeter wave devices. Given that blockage is expected to have a serious
effect on the link budget of commercial/cellular systems, a number of system level design questions
become pertinent. It is important to understand how different user hand grips/holdings can affect
signal strength behavior (perhaps generate some parametric models to capture these effects), how
blockage plays into network densification questions, and the role of mitigation mechanisms such as
multi-panel, multi-beam and cooperative schemes. Also, important to understand is the cost-power-performance
tradeoffs in the use of multiple antenna modules at millimeter wave frequencies~\cite{vasanth_tcom2019}.
Yet another broad question of interest is the tuning of hand phantoms to match true hand holding
results.

%{\vspace{-0.05in}}
\bibliographystyle{IEEEbib}
\bibliography{newrefsx2}

\end{document}